\documentclass[10pt,prd,preprintnumbers,longbibliography,floatfix,aps,nofootinbib,showpacs,twocolumn,superscriptaddress,noshowpacs]{revtex4}
\usepackage{epsfig}
\usepackage{url}
\usepackage{hyperref}
\usepackage[normalem]{ulem} 
\usepackage{latexsym}
\usepackage{epsfig}
\usepackage{amsmath}
\usepackage{amssymb}
\usepackage{wasysym}
\usepackage{graphicx}
\usepackage{verbatim}
\usepackage{enumerate,mdwlist}
\usepackage[titletoc]{appendix}
\usepackage{amsfonts}
\usepackage{tikz} 
\usetikzlibrary{calc}
\usetikzlibrary{mindmap,trees,shadows}
\usepackage{csquotes}
\tikzset{
  invisible/.style={opacity=0},
  dark/.style={opacity=0.4},
}
\usepackage{pgfplots}
\usepackage[export]{adjustbox}
\usepackage[normalem]{ulem}

\newcommand{\lp}{\left(}
\newcommand{\rp}{\right)}
\newcommand{\lb}{\left[}
\newcommand{\rb}{\right]}

\newcommand{\ba}{\begin{eqnarray}}
\newcommand{\ea}{\end{eqnarray}}
\newcommand{\be}{\begin{equation}}
\newcommand{\ee}{\end{equation}}

\newcommand{\al}{\alpha}

\newcommand{\Lag}{\mathcal{L}}
\newcommand{\A}{\mathcal{A}}

\newcommand{\aT}{\alpha_{_T}}

\newcommand{\lGW}{\lambda_{\mathrm{gw}}} 
\newcommand{\gb}{g^{_\text{B}}}
\newcommand{\hmn}{h_{\mu\nu}}
\newcommand{\thmn}{\bar{h}_{\mu\nu}}
\newcommand{\gmn}{g_{\mu\nu}}
\newcommand{\gbmn}{\gb_{\mu\nu}}

\newcommand{\ud}[2]{^{#1}_{\phantom{#1} #2}}
\newcommand{\du}[2]{_{#1}^{\phantom{#1} #2}}

\definecolor{grey}{rgb}{0.4,0.4,0.4}
\definecolor{dullmagenta}{rgb}{0.4,0,0.4}
\definecolor{darkblue}{rgb}{0,0,0.4}
\definecolor{midblue}{rgb}{0,0,0.5}
\definecolor{midred}{rgb}{0.5,0,0}
\definecolor{orange}{rgb}{1,0.5,0}
\definecolor{lightbrown}{rgb}{0.75,0.5,0.25}
\definecolor{tan}{cmyk}{0.14,0.42,0.56,0}
\definecolor{djunglegreen}{cmyk}{0.99,0,0.52,0}
\definecolor{lightgreen}{rgb}{0,1,0}
\definecolor{olivegreen}{cmyk}{0.64,0,0.95,0.40}
\definecolor{midgreen}{rgb}{0.0,0.675,0.0}
\definecolor{darkgreen}{rgb}{0,0.5,0}
\usepackage[all]{xy} 
\usepackage{amsfonts}
\preprint{IFT-UAM-CSIC-18-83, CERN-TH-2018-172}
\begin{document} 
\title{Dark Energy in light of Multi-Messenger Gravitational-Wave astronomy}
\author{Jose Mar\'ia Ezquiaga}
\email{jose.ezquiaga@uam.es}
\affiliation{Instituto de F\'isica Te\'orica UAM/CSIC, Universidad Aut\'onoma de Madrid, \\ 
C/ Nicol\'as Cabrera 13-15, Cantoblanco, Madrid 28049, Spain}
\author{Miguel Zumalac\'arregui}
\email{miguelzuma@berkeley.edu}
\affiliation{Berkeley Center for Cosmological Physics, LBNL and University of California at Berkeley, \\
Berkeley, California 94720, USA}
\affiliation{Institut de Physique Th\' eorique, Universit\'e  Paris Saclay 
CEA, CNRS, 91191 Gif-sur-Yvette, France}

\begin{abstract}
Gravitational waves (GWs) provide a new tool to probe the nature of dark energy (DE) and the fundamental properties of gravity. We review the different ways in which GWs can be used to test gravity and models for late-time cosmic acceleration.
Lagrangian-based gravitational theories beyond general relativity (GR) are classified into those breaking fundamental assumptions, containing additional fields and massive graviton(s).
In addition to Lagrangian based theories we present the effective theory of DE and the $\mu$-$\Sigma$ parametrization as general descriptions of cosmological gravity. 
Multi-messenger GW detections can be used to measure the cosmological expansion (standard sirens), providing an independent test of the DE equation of state and measuring the Hubble parameter.
Several key tests of gravity involve the cosmological propagation of GWs, including anomalous GW speed, massive graviton excitations, Lorentz violating dispersion relation, modified GW luminosity distance and additional polarizations, which may also induce GW oscillations. 
We summarize present constraints and their impact on DE models, including those arising from the binary neutron star merger GW170817. 
Upgrades of LIGO-Virgo detectors to design sensitivity and the next generation facilities such as LISA or Einstein Telescope will significantly improve these constraints in the next two decades.
\end{abstract}
\date{\today}
\maketitle
\tableofcontents

\begin{figure*}
\centering 
 \includegraphics[width=0.9\textwidth]{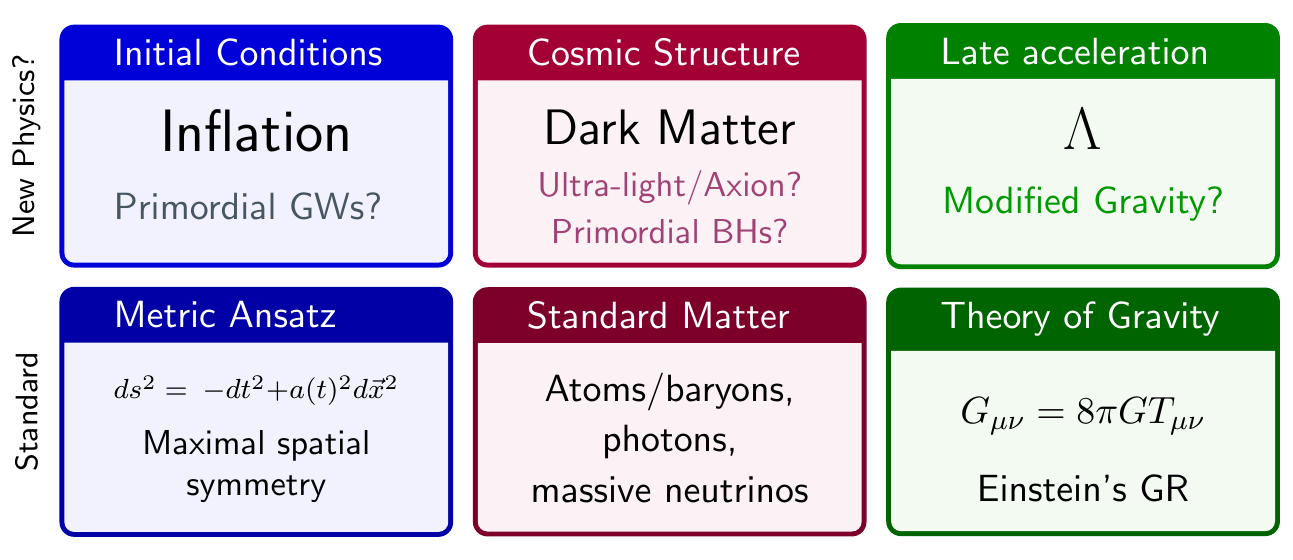}
 \vspace{-10pt}
 \caption{Ingredients of the Standard Model of Cosmology and their possible connection with new physics.}
 \label{fig:SMcosmo}
\end{figure*}

\section{Introduction}

The Standard Model of Cosmology (or $\Lambda$CDM) stands as a robust description of our universe. It is based on the theory of General Relativity (GR), which dictates the long-range gravitational interactions, together with the Cosmological Principle, which describes the geometry as homogeneous and isotropic on large scales. Standard matter (baryons, photons, neutrinos...) represents only a small fraction of the energy budget of the universe. The main ingredient is dark energy (DE), an unknown substance causing the late time acceleration. The other major component is dark matter (DM), an undetected constituent that seeds cosmic structures. The last piece of the Standard Model (SM) of Cosmology are the initial conditions, which are thought to be set by an early period of quasi-exponential expansion known as inflation. Despite the observational success of this model \cite{Aghanim:2018eyx}, it remains as a puzzle the fundamental origin of each piece, which could be associated to new physics (see Fig. \ref{fig:SMcosmo} for a summary of the different ingredients). 

In the SM of Cosmology, the current accelerated expansion is explained by a constant energy density acting as a perfect fluid with negative pressure. Such a cosmological constant (CC) term is perfectly consistent with present observations but notoriously disagrees with theoretical expectations for the vacuum energy \cite{Weinberg:1988cp,Martin:2012bt}. If this energy density is let to evolve in time, one naturally arrives to a dynamical description of DE sourced by a cosmological scalar field \cite{Copeland:2006wr}. If this field is now allowed to interact (non-minimally) with gravity, the possibilities to describe the cosmic expansion escalate \cite{Clifton:2011jh}. Alternatives to $\Lambda$CDM offer the possibility to alleviate some of its tensions. For instance, DE models with an effective equation of state more negative than the cosmological constant could ease the tension between the local measurement of the Hubble constant and the inferred value from the cosmic microwave background (CMB). Exploring the largest scales with galaxy surveys like Euclid or LSST will help us understanding the expansion history of the universe and will provide new insights about gravity.

Gravity can be tested at different scales and regimes. Classical tests of gravity range from laboratory experiments to Solar System distances, and cover gravity in its weak field regime \cite{Will:2014kxa}. Astrophysical observations provide new avenues to improve these tests \cite{Berti:2015itd}. Pulsars in particular can be especially constraining, for instance with the recent observations of a triple stellar system \cite{Archibald:2018oxs}. Tests in a much stronger regime have been performed tracking stellar orbits around the galactic center \cite{Hees:2017aal}. Altogether, these observations severely constrain modifications of GR. Theories beyond Einstein's theory should thus resemble GR at small scales, e.g. hiding fifth forces with screening mechanisms \cite{Brax:2013ida,Joyce:2014kja}. At large scales, however, present constraints are considerably weaker. Combining different probes could be crucial to set an observational program to test gravity from cosmology \cite{Weinberg:2012es}.

Gravitational wave (GW) astronomy offers the possibility to test gravity both in the strong regime and at large scales. So far there have been six individual detections, five binary black-holes (BBH) \cite{Abbott:2016blz,Abbott:2016nmj,Abbott:2017vtc,Abbott:2017gyy,Abbott:2017oio} and one binary neutron star (BNS) \cite{TheLIGOScientific:2017qsa}. No GW background \cite{Abbott:2017xzg}, periodic source \cite{Abbott:2018bwn} or long-duration transient \cite{Abbott:2017muc} have been detected. 

GWs could be critical in resolving the open problems of the SM of Cosmology. For instance, (non) observations of cosmological backgrounds of primordial GWs test inflation. BBHs events teach us about the population of BHs, which constrains their possible contribution to DM and their possible primordial origin \cite{Sasaki:2018dmp}. Moreover, if DM is described by ultra-light bosons or axions, it could resonate with pulsars \cite{Blas:2016ddr} or form clouds around BHs observable with GWs \cite{Arvanitaki:2016qwi}. Finally, BNS with an associated counterpart such as GW170817 \cite{Monitor:2017mdv,GBM:2017lvd} become standard sirens \cite{Abbott:2017xzu} and allow to probe DE. In this review we will focus on this last case, exploring the possibilities of multi-messenger GW astronomy to probe the nature of DE and the fundamental properties of gravity (see a schematic timeline of present and future facilities in Fig.~\ref{fig:MMtimeline}).

\subsection{Summary for the busy reader}
 
\begin{figure*}[t]
\includegraphics[width=\textwidth]{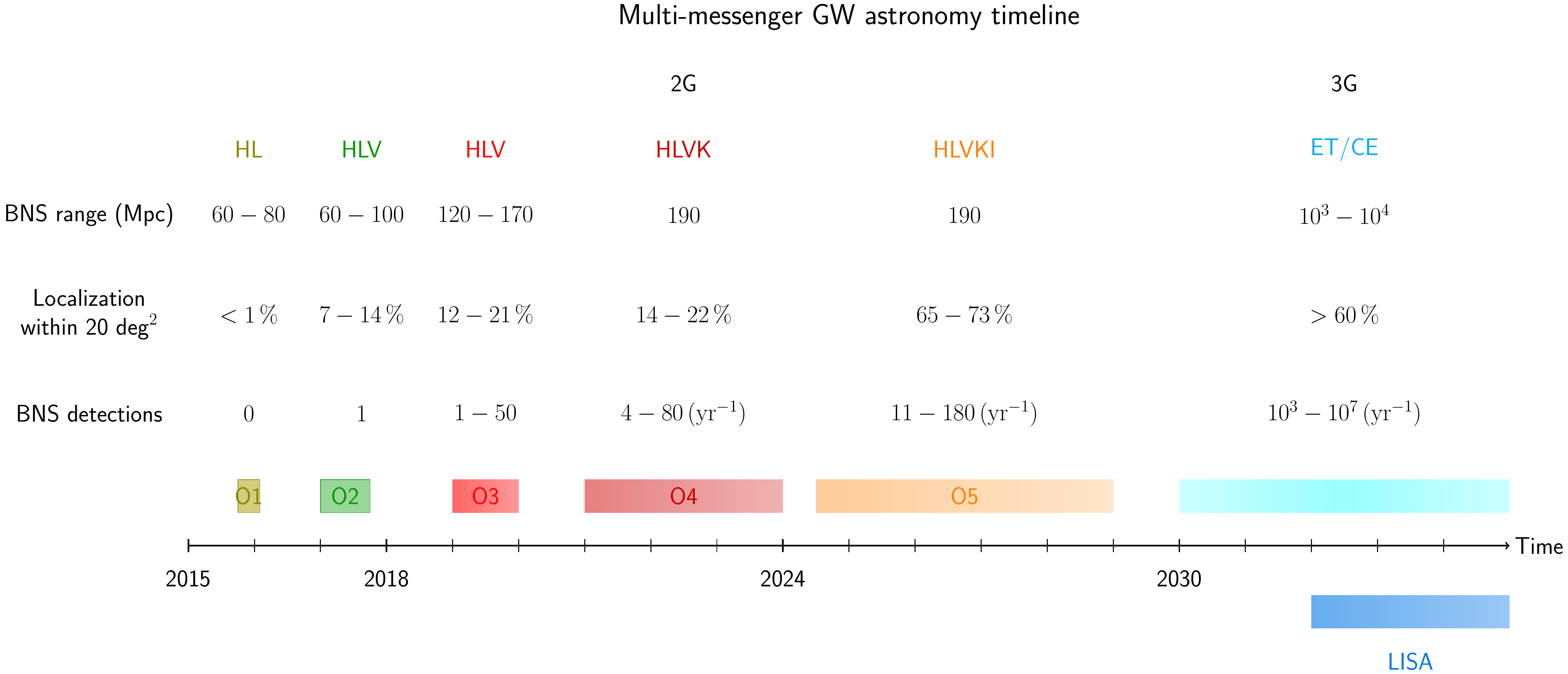}
\vspace{-15pt}
\caption{Schematic multi-messenger GW astronomy timeline. The binary neutron star (BNS) rate, the localization area in the sky, and the number of BNS detections are given for past and future observation runs. Second generation (2G) ground-based detectors organize in five runs (O1-O5) with different number of detectors online (from 2 to 5) \cite{Aasi:2013wya}. The nomenclature used is H=Hanford, L=Livingston, V=Virgo, K=KAGRA, I=IndIGO. Third generation (3G) detectors projected are Einstein Telescope (ET) \cite{Sathyaprakash:2012jk} or Cosmic Explorer (CE) \cite{Evans:2016mbw}. The localization in 3G depends on the network of detectors which is still uncertain \cite{Mills:2017urp}. For reference, we include the timeline space-based detector LISA \cite{AmaroSeoane:2012km}. The reader should note that this numbers correspond to present expectations. For more details we refer to Sec. \ref{sec:Detectors}.} \label{fig:MMtimeline}
\end{figure*}

Dark energy is the major component of the universe and yet its nature escapes our present understanding. Beyond the cosmological constant paradigm, a plethora of alternative theories of gravity has been proposed to explain the current cosmic acceleration (see Fig. \ref{fig:theory_roadmap} for a roadmap of possible modifications of gravity). We present an overview of the landscape of theories in Sec. \ref{sec:TheoryGravityDE}, as well as a summary of the different approaches to cosmological gravity (see Fig.~\ref{fig:effective_gravity} for a schematic diagram).
 
Gravitational wave astronomy opens new possibilities to probe gravity and DE. For readers unfamiliar with the basics of GWs, we provide a short introduction in Sec. \ref{sec:BasicsGW}. For the purpose of cosmology, the most promising GW events are those that can be observed by other messengers (either EM waves or neutrinos). There are four main tests one can do with multi-messenger GW events:
 \begin{itemize}
 \item \emph{Standard sirens (Sec. \ref{sec:StandardSirens}):} the amplitude of GWs is inversely proportional to its luminosity distance. If a counterpart of the GW is observed, a redshift measurement of the source is possible and the cosmic expansion history can be constrained. For close by sources, only the Hubble constant is measured. Future standard sirens measurements could help resolving the present tension in $H_0$ (see Fig. \ref{fig:Hubbletension}).
 
 \item \emph{GW speed (Sec. \ref{sec:GWspeed}):} the propagation speed of GWs follows from the dispersion relation. Once the location of a GW event is known, it is possible to compare the speed of GWs with respect to the speed of light. Many alternative gravity theories predict that GWs propagate at a different speed either by modifying the effective metric in which GWs propagate, by inducing a mass for the graviton or by introducing higher order terms in the dispersion relation.
 
 \item \emph{GW damping (Sec. \ref{sec:GWdamping}):} modified gravity interactions can also alter the amplitude of GWs. In addition to the cosmic expansion, effective friction terms can damp GWs. This introduces an inequality between the GW and the EM luminosity distance that can be tested.
 
 \item \emph{Additional polarizations (Sec. \ref{sec:GWPolarization}):} in alternative theories of gravity, there could be additional modes propagating. These extra polarizations could be directly tested if the source is localized and there is a network of detectors online. Moreover, these modes could mix with the tensor perturbations leading, for instance, to GW oscillations.
 
\end{itemize}

In this review we aim at summarizing current bounds on gravity theories and dark energy models from the first multi-messenger GW detection, GW170817. Up to date, the most constraining test is the GW speed. We also survey the prospects of different multi-messenger tests with future detectors. Significant improvements can be achieved in probing the GW Hubble diagram with an increasing number of events. A schematic timeline of multi-messenger GW astronomy is presented in Fig.~\ref{fig:MMtimeline} (the reader should be aware that expectations far in the future are very preliminary). The theoretical implications of present and future observations are discussed in  Sec. \ref{sec:TheoreticalImplications}. We close the work in Sec. \ref{sec:Conclusions} with an outlook of prospects and challenges of multi-messenger GW tests of gravity and DE.

\section{Theories of gravity and Dark Energy}
\label{sec:TheoryGravityDE}

\begin{figure*}[t]
\includegraphics[width=\textwidth]{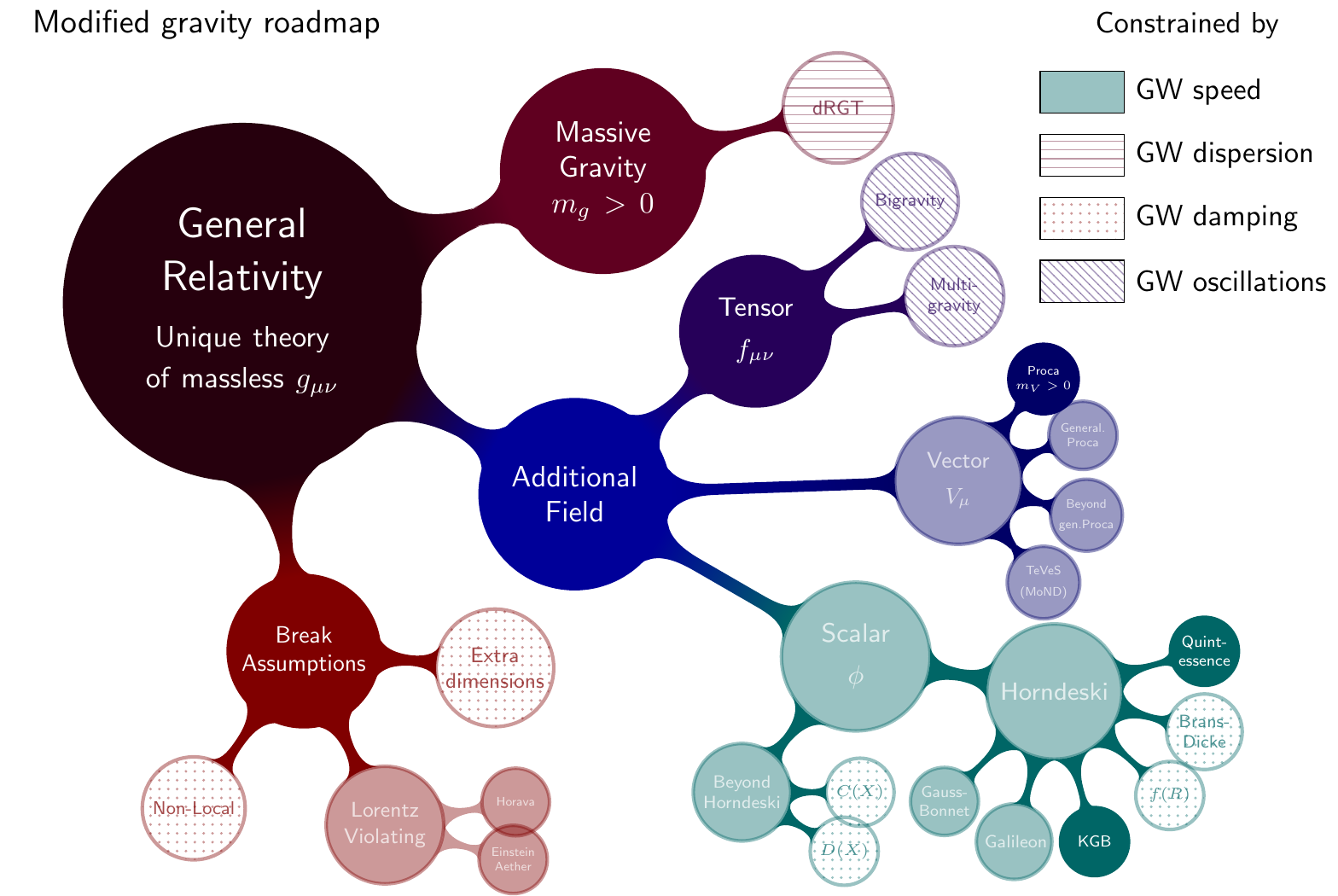}
\caption{Modified gravity roadmap summarizing the possible extensions of GR described in Sec.~\ref{sec:TheoryGravityDE}. The main gravitational wave (GW) test of each theory is highlighted. For details in the different tests see the discussion in Sec. \ref{sec:GWspeed} (GW speed and dispersion), \ref{sec:GWdamping} (GW damping) and \ref{sec:GWPolarization} (GW oscillations). Theories constrained by the GW speed and GW oscillations can also be tested with GW damping and GW dispersion respectively. Note in addition that many theories fall under different categories of this classification (see text in Sec.~\ref{sec:covariant_theories}).
} 
\label{fig:theory_roadmap}
\end{figure*}

The quest to test gravity and find alternatives to the cosmological constant has produced many theories beyond Einstein's General Relativity (GR) and other descriptions of gravity on cosmological scales. 
We will classify the different means to modify Einstein's theory and review their status as descriptions of cosmic acceleration.
Then we will review other general approaches to describe gravity on cosmological scales, namely through the effective theory of dark energy and phenomenological parameterizations of the gravitational potentials.
The landscape of alternative theories is summarized in Fig. \ref{fig:theory_roadmap} and the approaches to cosmological gravity are schematically described in Fig. \ref{fig:effective_gravity}.

\subsection{Theories of Gravity} \label{sec:covariant_theories}

The action-based approach to modify gravity is based on generalizing the Einstein-Hilbert action
\begin{equation}\label{eq:GR_action}
 S_{GR} = \int d^4 x \sqrt{-g} \frac{R[g_{\mu\nu}]}{16\pi G} + S_m[g_{\mu\nu},\cdots]\,,
\end{equation}
where $G$ is Newton's constant and $S_m$ denotes the action of matter, universally and minimally coupled to the metric $g_{\mu\nu}$. Variation of the action (\ref{eq:GR_action}) with respect to the metric leads to Einstein's field equations
\begin{equation} \label{eq:EinsteinEquation}
 G_{\mu\nu} \equiv R_{\mu\nu} - \frac{1}{2}R g_{\mu\nu} = 8\pi G T_{\mu\nu}\,,
\end{equation}
where $R_{\mu\nu}$ is the Ricci tensor, $R\equiv g^{\mu\nu}R_{\mu\nu}$ the Ricci scalar and $T_{\mu\nu}=\frac{-2}{\sqrt{-g}}\frac{\delta S_m}{\delta g^{\mu\nu}}$ is the matter energy-momentum tensor.
Einstein's equations can be used to obtain solutions for the space-time ($g_{\mu\nu}$) given the matter content ($T_{\mu\nu}$) in any physical situation, including cosmological solutions relevant to study dark energy.

The structure of gravitational theories is severely restricted and several results can be used to prove the uniqueness of General Relativity under quite broad assumptions. 
Weinberg's theorems restrict possible infrared (low energy) interactions of massless, Lorentz invariant particles, which for spin-2 lead unavoidably to the equivalence principle \cite{Weinberg:1964ew} and the derivation of Einstein's equations \cite{Weinberg:1965rz}.\footnote{In addition to GR, there is another theory for massless, spin-2 fields in 4D, Unimodular Gravity, which is invariant under diffeomorphisms preserving the 4D volume element \cite{vanderBij:1981ym}.}
At the classical level, the results of Lovelock imply that the Einstein-Hilbert action is unique in 4D~\cite{Lovelock:1971yv,Lovelock:1972vz}.

According to the above results, alternative theories of gravity can be classified into those that
\begin{itemize}
 \item Break the fundamental assumptions.
 \item Include additional fields.
  \item Make the graviton massive. 
\end{itemize}
Note that those descriptions are not exclusive, and many theories fall within several categories. For instance: bimetric gravity has an additional field (tensor) and contains a massive graviton, Einstein-Aether is both Lorentz-violating and includes a vector field, TeVeS has a scalar in addition to a vector, and many extra-dimensional models can be described in terms of additional fields in certain limits. Also, when referring to massive gravitons, we will be considering only classical spin-2 fields.

\subsubsection{Breaking fundamental assumptions}

The theorems that fix the structure of General Relativity assume a four dimensional pseudo-Riemannian manifold and local interactions satisfying Lorentz invariance. Any departure from these principles offers a way to construct modified theories of gravity.%
\footnote{A class of GR extensions include additional geometric elements like torsion or non-metricity. These elements can be viewed as either breaking the fundamental assumptions or including additional fields.}

\paragraph{Extra dimensions:}
Additional spatial dimensions allow the inclusion of new operators constructed only from the metric tensor. The canonical example are Lovelock invariants \cite{Lovelock:1971yv}, such as the Gauss-Bonnet term (a topological term in 4 dimensions which does not contribute to the equations of motion). 
The lack of observation of extra dimensions requires some mechanism to hide them. One example is compactification, when extra dimensions are sufficiently small that they are not accessible to experimental tests \cite{Bailin:1987jd,Overduin:1998pn}. 
A radically opposite view consist on Braneworld constructions, in which the standard model fields live in a 3+1 dimensional brane, embedded in the higher dimensional space \cite{ArkaniHamed:1998rs,Antoniadis:1998ig,Randall:1999ee}.
The Dvali-Gabadadze-Porrati (DGP) model \cite{Dvali:2000hr,Nicolis:2004qq} is one such construction in which self-accelerating solutions\footnote{Self-accelerating solutions are those in which there is a late time acceleration without a cosmological constant ($\Lambda=0$).} can be obtained. However, this branch of solutions is plagued by a ghost instability. The 4D effective theory can avoid this problem and was the origin of Galileon gravity \cite{Nicolis:2008in}.

\paragraph{Lorentz Invariance Violation}
Gravity can be extended by breaking Lorentz invariance. 
In many of these alternatives a preferred time direction emerges spontaneously breaking Lorentz symmetry (see \cite{Blas:2014aca} for a review).
Ho\v{r}ava gravity \cite{Horava:2009uw} implements Lorentz violation through a preferred foliation of space-time, with the attractive property that Lorentz symmetry can be recovered at low energies (see \cite{Blas:2009qj,Blas:2009yd,Sotiriou:2009gy,Sotiriou:2009bx,Sotiriou:2010wn} for extensions/variants) and leading to a power-counting renormalizable theory of gravity.
Another class of Lorentz-violating theories is Einstein-Aether, in which a vector field with constant norm introduces a preferred direction \cite{Jacobson:2008aj}. The special case of Einstein-Aether theories in which the vector field is the gradient of a scalar is known as Khronometric \cite{Blas:2010hb}. Khronometric theories describe the low-energy limit of some extension of Ho\v{r}ava-gravity, linking the two frameworks \cite{Jacobson:2010mx}. 
These ideas have been studied as cosmological scenarios \cite{Audren:2013dwa,Audren:2014hza}.

\paragraph{Non-local theories}
Non-local theories include inverse powers of the Laplacian operator. 
These models can involve general functions  (e.g. $R\cdot f(\Box^{-1}R)$) \cite{Deser:2007jk,Koivisto:2008dh} or be linear (e.g. $R\frac{m^2}{\Box^2} R$) \cite{Jaccard:2013gla}.
The latter class of models lead to phantom dark energy \cite{Maggiore:2013mea,Maggiore:2014sia} and are compatible with cosmological observations \cite{Dirian:2014bma} (see \cite{Maggiore:2014sia} for a review). However, their viability on the solar system is disputed due to the time evolution of the effective degrees of freedom and the lack of a screening mechanism \cite{Barreira:2014kra}.
Non-local interactions have been also proposed as a means to improve the ultra-violet behavior of gravity \cite{Biswas:2011ar,Modesto:2011kw,Calcagni:2014vxa}.
Non-local models are constructed using the Ricci scalar, since non-local terms involving contractions of the Ricci tensor give rise to cosmological instabilities \cite{Ferreira:2013tqn,Nersisyan:2016jta}.

\subsubsection{Additional fields}

Gravity can be extended by the inclusion of additional fields that interact directly with the metric. These theories will vary by the type of field (scalar, vector, tensor) and the interaction with gravity it has. Theories with additional tensors (bigravity and multigravity) are extensions of massive gravity and will be described in Sec. \ref{sec:massive_gr}.
We will assume a minimal universal coupling of matter to the metric.
For a very complete review of gravity theories containing additional fields, see Ref. \cite{Heisenberg:2018vsk}.

\paragraph{Scalar field}

A scalar is the simplest field by which gravity can be extended. 
Scalars do not have a preferred orientation and thus a macroscopic, classical state can exist in the universe without affecting the isotropy of the space-time if it depends only on time. 
Moreover, a potential term can mimic a cosmological constant very closely in the limit in which the field is varying very slowly (e.g. if the potential is very flat), which is the foundation of the simplest single-field inflation and dark energy models (quintessence).
Scalar fields may also arise in effective descriptions of fundamental theories belonging to other categories, such as braneworld constructions \cite{deRham:2010eu,Goon:2011uw,Koivisto:2013fta}. These properties had led to a proliferation of scalar-based models to describe accelerating cosmologies, both in the context of inflation and dark energy. 

Recent efforts to study scalar-tensor theories have led to a classification based on the highest-order derivatives of the additional field present in the action and the equations of motion, with three generations of theories
\begin{enumerate}
 \item Old-school scalar tensor theories: 1$^{\rm st}$ order derivatives in the action, $2^{\rm nd}$ order in equations.
 \item Horndeski theories \cite{Horndeski:1974wa}: 2$^{\rm nd}$ order derivatives in the action and $2^{\rm nd}$ order in equations.
 \item Beyond Horndeski: 2$^{\rm nd}$ order derivatives in the action and higher order in equations.
\end{enumerate}
The classification is motivated by Ostrogradski's theorem, which states that theories with second and higher (time) derivatives in the action generically introduce unstable degrees of freedom \cite{Ostrogradski,Woodard:2015zca}.
While most physical theories belong to the first class, known loopholes to Ostrogradski's theorem exits, for instance in effective or non-local theories (in which the ghost degrees of freedom are removed) \cite{Simon:1990ic} or when the theory is degenerate (that is, the inversion to canonical variables is not possible).
The degeneracy condition is automatically satisfied if the equations of motion are second order, but that is not strictly necessary (different conditions appear when there are additional degrees of freedom \cite{Motohashi:2016ftl}).%
\footnote{Scalar-tensor theories can be reformulated in terms of differential forms in which the second order equations follow naturally from the antisymmetry of this language \cite{Ezquiaga:2016nqo}. This approach can be generalized to gravity theories with additional vector and tensor fields as well \cite{Ezquiaga:2017ner}.}
Known viable beyond Horndeski theories are known as Degenerate Higher Order Scalar Tensor (DHOST) \cite{Langlois:2015cwa}, which have second derivatives in the action (higher derivatives in the equations), but recently toy models with higher derivatives in the action have been proposed \cite{Motohashi:2018pxg}.

Old-school scalar-tensor theories contain at most first derivatives of the scalar in the action. They can be seen as a generalization of the Jordan-Brans-Dicke theory of gravity \cite{Brans:1961sx}
\begin{equation}\label{eq:lag_brans_dicke}
S=\int d^4x \sqrt{-g}\frac{M_{\rm Pl}^2}{2}\left[\omega(\phi) R-K(X,\phi)\right]+ S_{m} \,, 
\end{equation}
where $X\equiv-\nabla^\nu\phi\nabla_\nu\phi/2$ is the canonical kinetic term of the scalar field. This theory includes GR ($\omega=1,K=\Lambda$),
quintessence ($\omega=1, K=X-V$) \cite{Wetterich:1987fm,Ratra:1987rm}, Brans-Dicke models \cite{Brans:1961sx}  ($\omega=\phi$, $K=\frac{\omega_{\rm BD}}{\phi} X - V(\phi)$),
k-\emph{essence} \cite{ArmendarizPicon:1999rj,ArmendarizPicon:2000ah} ($\omega=1$, $K=K(\phi,X)$).
Archetypal modified-gravity models such as \emph{$f(R)$} \cite{Carroll:2003wy,Hu:2007nk,Sotiriou:2008rp} are equivalent to instances of these theories \cite{DeFelice:2010aj}. 
Chameleons \cite{Khoury:2003aq} 
and symmetrons \cite{Hinterbichler:2010es} 
also belong to this class of theories (see \cite{Burrage:2016bwy} for a review).
Certain freedom exists in writing the theory due to the possibility of rescaling the metric $g_{\mu\nu}\to \bar g_{\mu\nu} = C(\phi)g_{\mu\nu}$ and redefining the scalar field, i.e. the Jordan frame in which the metric is minimally coupled (\ref{eq:lag_brans_dicke}) and the Einstein frame in which $\omega$ is constant but matter is explicitly coupled to the scalar \cite{Flanagan:2004bz}. 
 Current cosmological observations constrain the Brans-Dicke parameter $\omega_{\rm BD}> 692$ (99\%) \cite{Avilez:2013dxa}.

Horndeski's theory contains the best understood examples of scalar-tensor theories. 
The Horndeski action encompasses all local, 4D Lorentz invariant actions whose metric and field variation leads to second order equations of motion \cite{Horndeski:1974wa} (Horndeski's theory is also known in the literature as Generalized Galileons \cite{Deffayet:2011gz,Kobayashi:2011nu}). 
Horndeski's action reads
\begin{eqnarray}\label{eq:L_horndeski}
S=\int d^4x \sqrt{-g}\sum_{i=2}^5{\cal L}_i[\phi,g_{\mu\nu}]+ S_{m} [\chi_i ,g_{\mu \nu}],
\end{eqnarray}
where we have assumed minimal and universal coupling to matter in $S_m$. The sum is over the four Lagrangians
\begin{eqnarray}
{\cal L}_2&=& K(X,\phi) ,  \label{eq:LH2} \\
{\cal L}_3&=&  -G_3(X,\phi) \Box\phi , \label{eq:LH3} \\
{\cal L}_4&=&   G_4(X,\phi)R+G_{4X}\left\{(\Box \phi)^2-\phi_{\mu\nu} \phi^{\mu\nu}\right\}  , \label{eq:LH4} \\
{\cal L}_5&=& G_5(X,\phi)G_{\mu\nu}\phi^{\mu\nu}
-\frac{1}{6}G_{5X}\big\{ (\Box\phi)^3
-3\phi^{\mu\nu}\phi_{\mu\nu}\Box\phi \nonumber \\ & & \qquad \qquad \qquad\qquad
+2\phi\du{\nu}{\mu} \phi\du{\alpha}{\nu}\phi\du{\mu}{\alpha}
\big\}  \,,
\label{eq:LH5}
\end{eqnarray}
where $K$ and $G_A$ are functions of $\phi$ and $X\equiv-\nabla^\nu\phi\nabla_\nu\phi/2$, and the subscripts $X$ and $\phi$ denote partial derivatives. 
Horndeski theories include all the generalized Jordan-Brans-Dicke type, plus new additions that involve second derivatives of the scalar at the level of the action. 
These include kinetic gravity braiding (KGB) ($K(X),G_3(X)$) \cite{Deffayet:2010qz,Kobayashi:2010cm,Pujolas:2011he},
covariant galileons ($K,G_3\propto X$, $G_4,G_5\propto X^2$) \cite{Nicolis:2008in,Deffayet:2009wt}, disformal \cite{Koivisto:2012za}
and Dirac-Born-Infeld gravity ($G_i \propto \sqrt{1+X/\Lambda_i^4}$) \cite{deRham:2010eu,Zumalacarregui:2012us}, 
Gauss-Bonnet couplings \cite{Ezquiaga:2016nqo} 
and models self-tuning the cosmological constant \cite{Charmousis:2011bf,Martin-Moruno:2015bda}.
Just as Brans-Dicke is invariant under rescalings of the metric, Horndeski theories are invariant under field-dependent disformal transformations $g_{\mu\nu}\to \bar g_{\mu\nu} = C(\phi)g_{\mu\nu} + D(\phi)\phi_{,\mu}\phi_{,\nu}$, which amount to a redefinition of the Horndeski functions $G_i$ (and the introduction of an explicit coupling to matter) \cite{Bettoni:2013diz}.

Theories beyond Horndeski have higher order equations of motion without including additional degrees of freedom. The first examples of these theories \cite{Zumalacarregui:2013pma} were related to GR by a metric redefinition involving derivatives of the scalar field \cite{Bekenstein:1992pj},
\begin{equation} \label{eq:disf_gen_transf}
 g_{\mu\nu}\to \bar g_{\mu\nu} = C(X,\phi) g_{\mu\nu} + D(X,\phi)\phi_{,\mu}\phi_{,\nu}\,,
\end{equation}
applied to the gravity sector. The simplest such beyond Horndeski theory emerged from the metric rescaling with derivative dependence $ C= \Omega^2(X,\phi), D=0$, and was dubbed \emph{kinetic conformal gravity} \cite{Zumalacarregui:2013pma}
\begin{equation}\label{eq:kin_conf_theory}
S_{C} = \int d^4 x\frac{\sqrt{-g}}{16\pi G} \lp \Omega^2 R + 6 \Omega_{,\al}\Omega^{,\al} \rp + S_\phi + S_M\,,
\end{equation}
where $S_\phi$ is an additional scalar field Lagrangian. One of the premises in constructing this type of theory was the existence of an inverse for the relation (\ref{eq:disf_gen_transf}), which can be studied through the Jacobian of the mapping \cite{Zumalacarregui:2013pma}. 
If this assumption is broken the resulting theory is \emph{mimetic gravity} \cite{Chamseddine:2013kea}, a gravitational alternative to dark matter.
Interestingly, the conformal relation between kinetic conformal gravity (\ref{eq:kin_conf_theory}) and GR ensures that this is one of the theories in which the speed of GWs is nontrivially equivalent to the speed of light \cite{Ezquiaga:2017ekz,Creminelli:2017sry}.

The best known beyond Horndeski theory is given by the Gleyzes-Langlois-Piazza-Vernizzi (GLPV) action \cite{Gleyzes:2014dya}, which consists of Horndeski plus the additional Lagrangian terms:
\begin{eqnarray}
\label{eq:glpv4}
{\cal L}_{4b}&=&   B_4(\phi,X)\epsilon\ud{\mu\nu\rho}{\sigma}\epsilon^{\mu^\prime\nu^\prime\rho^\prime\sigma}\phi_{\mu}\phi_{\mu^\prime}\phi_{\nu\nu^\prime}\phi_{\rho\rho^\prime} \,,  \\
{\cal L}_{5b}&=&
\label{eq:glpv5}
B_5(\phi,X)\epsilon^{\mu\nu\rho\sigma}\epsilon^{\mu^\prime\nu^\prime\rho^\prime\sigma}\phi_{\mu}\phi_{\mu^\prime}\phi_{\nu\nu^\prime}\phi_{\rho\rho^\prime}\phi_{\sigma\sigma^\prime}\,. 
\end{eqnarray}
Horndeski and GLPV Lagrangians of the same order, i.e. $\Lag_4+\Lag_{4b}$ (\ref{eq:LH4}+\ref{eq:glpv4}) or $\Lag_5+\Lag_{5b}$ (\ref{eq:LH5}+\ref{eq:glpv5}), can be mapped to Horndeski via $g_{\mu\nu}\to \hat g_{\mu\nu} = C(\phi)g_{\mu\nu}+D(X,\phi)\phi_\mu \phi_\nu$ showing the viability of these combinations \cite{Gleyzes:2014dya,Gleyzes:2014qga}.
For generic combinations of Horndeski and GLPV, viability arguments were first based on a special gauge (unitary gauge) that assumed that the scalar field derivative $\phi_\mu$ is timelike. 
Subsequent analyses eventually lead to covariant techniques to study the degeneracy conditions \cite{Langlois:2015cwa} (see \cite{Deffayet:2015qwa} for earlier criticism). 
These techniques later showed that not all Horndeski and GLPV combinations met the degeneracy condition on a covariant level  \cite{Crisostomi:2016tcp}.

The study of degeneracy conditions for scalar-tensor theories ultimately led to the \textit{degenerate higher-order scalar-tensor} (DHOST) \cite{Langlois:2015cwa} paradigm classification of theories with the right number of degrees of freedom (also known as Extended Scalar-Tensor or EST) \cite{Crisostomi:2016czh}.
DHOST theories include cases beyond conformal kinetic gravity (\ref{eq:kin_conf_theory}) and GLPV theories (\ref{eq:glpv4},\ref{eq:glpv5}). 
DHOST theories are invariant under general disformal transformations (\ref{eq:disf_gen_transf}), which can in turn be used to classify them \cite{Achour:2016rkg} (see also \cite{Crisostomi:2017aim}).
DHOST theories have been fully identified including terms with up to cubic second-field derivatives in the action, e.g. $\sim (\Box\phi)^3$ \cite{BenAchour:2016fzp}.
Demanding the existence of a Poisson-like equation for the gravitational potential restricts the space of DHOST theories to those that are related to Horndeski via disformal transformations (\ref{eq:disf_gen_transf})~\cite{Langlois:2017mxy}.

\paragraph{Vector field}

Theories with vector fields have been proposed as modifications to GR and in the context of dark energy. A background vector field does not satisfy the isotropy requirements of the cosmological background, unless it points in the time direction and only depends on time $A_\mu =( A_0(t),0,0,0)$.
Isotropy can also happen on average, if a vector with a space-like projection oscillates much faster than the Hubble time \cite{Cembranos:2012kk}.
In that case the background is isotropic on average but the perturbations (including gravitational waves) inherit a residual anisotropy \cite{Cembranos:2016ugq}.
Finally, theories with multiple vectors can satisfy isotropy, for instance, if they are in a triad configuration $A_\mu^a=A(t)\delta_\mu^a$~\cite{ArmendarizPicon:2004pm}.\footnote{Technically speaking, multiple vectors can lead to isotropic solutions if they have an internal symmetry that together with the broken space-time symmetries leaves a residual $ISO(3)$ \cite{BeltranJimenez:2018ymu}. For the case of the triad, the symmetry group is $SO(3)$.}
A large number of vectors can also lead to statistical isotropy (e.g. if the orientations are random) \cite{Golovnev:2008cf}. 
The kinetic term for a vector field, $F_{\mu\nu}F^{\mu\nu}$, is defined by the gauge invariant field strength $F_{\mu\nu}=\partial_{\mu}A_{\nu}-\partial_{\nu}A_{\mu}$ and the addition of a mass term $m^2 A_\mu^2$ is known as Proca theory \cite{Proca:1900nv}.

Proca theories have been generalized to include explicit gravitational interactions of a massive vector field \cite{Tasinato:2014eka,Heisenberg:2014rta,Allys:2015sht,Jimenez:2016isa}. 
The vector field Lagrangian is built so that precisely one extra (longitudinal) scalar mode propagates in addition to the two usual Maxwell-like transverse polarizations. 
Its full generalization contains terms with direct couplings between the vector and space-time curvature, whose structure closely resembles those of Horndeski's theory (\ref{eq:LH4},\ref{eq:LH5}).
In analogy to beyond Horndeski, there are also beyond generalized Proca interactions \cite{Heisenberg:2016eld,Kimura:2016rzw}.
Further extensions to multiple vector fields known as generalized multi-Proca/Yang-Mills theories are able to incorporate new couplings \cite{Allys:2016kbq} and configurations \cite{Jimenez:2016upj}, e.g. the extended triad $A_\mu^a=\phi^a\delta^0_\mu+A(t)\delta_\mu^a$, as do theories with a vector and a scalar (Scalar-Vector-Tensor) \cite{Heisenberg:2018acv}.
For more details about these theories we recommend Ref.~\cite{Heisenberg:2018vsk}.

An iconic theory containing a vector is the Tensor-Vector-Scalar (TeVeS) theory by Bekenstein \cite{Bekenstein:2004ne}. 
TeVeS emerged as a relativistic theory able to describe Modified Newtonian Dynamics (MOND), and thus as an alternative to dark matter.
For an overview of field-theoretical aspects of TeVeS and related theories, including other  relativistic MOND candidates, see Ref.  \cite{Bruneton:2007si}.
TeVeS theory introduces several non-minimal ingredients. In addition to the gravitational metric $\tilde g_{\mu\nu}$ matter is minimally coupled to an effective metric 
\begin{equation}
 g_{\mu\nu} = e^{-2\phi}\tilde g_{\mu\nu} - 2\sinh(2\phi)A_\mu A_\nu \,,
\end{equation}
which generalizes the scalar disformal relation (\ref{eq:disf_gen_transf}), incorporating the vector. 
Here $\tilde g_{\mu\nu}$ is the gravitational metric, $\phi$ is the scalar. The vector $A_\mu$ is enforced to be time-like and normalized with respect to the gravitational metric $\tilde g^{\mu\nu}A_\mu A_\nu = -1$. 
TeVeS has a very rich phenomenology, including effects in GW propagation \cite{Sagi:2010ei}.
At the level of cosmology it is partially able to mimic DM, although the oscillations of the fields make it hard for the theory to reproduce the peaks in the CMB \cite{Skordis:2005xk,Bourliot:2006ig,Skordis:2009bf}.
 
\subsubsection{Massive gravity and tensor fields}\label{sec:massive_gr}

Giving a mass to the graviton is another means to extend GR, with gravity mediated by a particle with mass $m_g$, spin $s=2$ and $2s+1=5$ polarization states 
(see \cite{deRham:2016nuf} for bounds on the graviton mass). 
Weinberg theorem on the structure of GR relies on the infrared properties of the interactions: a mass term changes this structure. 
Despite this clear loophole, constructing a self-consistent theory of massive gravity, free of pathologies and with the right number of degrees of freedom proved an extremely hard endeavor that took nearly 70 years to complete. 
The linear theory of massive gravity was formulated in 1939 by Fierz \& Pauli \cite{Fierz:1939ix} as linearized GR plus a mass term
\begin{equation}\label{eq:L_fierz_pauli}
 S_{\rm FP} = \int d^4x \; m_g^2\left(h^{\mu\nu}h_{\mu\nu} - (\eta^{\mu\nu}h_{\mu\nu})^2\right)\,.
\end{equation}
It was later found that Fierz-Pauli theory was discontinuous and gave different results from GR in the limit $m_g\to0$ (vDVZ discontinuity) \cite{vanDam:1970vg,Zakharov:1970cc}. 
The discrepancy is due to the longitudinal polarization of the graviton (the helicity-zero mode) not decoupling in that limit.
Considering non-linear interactions solved the apparent discontinuity by hiding the helicity-zero mode, which is strongly coupled in regions surrounding massive bodies and effectively decouples, recovering the GR predictions when $m_g\to 0$ \cite{Vainshtein:1972sx}.
Despite this progress, massive gravity had another important flaw: all theories seemed to have an additional mode (known as Boulware-Deser (BD) ghost) that renders the theory unstable \cite{Boulware:1973my,Creminelli:2005qk}.

 \paragraph{Ghost Free Massive Gravity}
The apparent difficulties were overcome in de Rham-Gabadadze-Tolley theory (dRGT) \cite{deRham:2010kj}, also known as ghost-free massive gravity (for current reviews on the theory see \cite{Hinterbichler:2011tt,deRham:2014zqa}).
dRGT is a ghost free theory propagating the 5 polarizations corresponding to a spin-2 massive particle, universally coupled to the energy-momentum tensor of matter (cf. Fig. \ref{fig:Polarizations}).
The ghost-free property was initially shown in the decoupling limit (in which the helicity-0 mode decouples from the other polarizations) and then in the full theory \cite{Hassan:2011hr,Hassan:2011ea} . 
The phenomenological deviations induced by massive gravity are primarily due to the helicity-0 mode. On small enough scales the Vainshtein mechanism \cite{Vainshtein:1972sx} (see \cite{Babichev:2013usa} for a review) effectively suppresses these interactions, leading to predictions very similar to GR on Solar System scales (however, new classes of solutions for black holes do exist, in addition to the usual ones \cite{Babichev:2015xha}).

Massive gravity may offer a solution to the accelerating universe. 
A heuristic argument is that the force mediated by the massive graviton has a finite range $V\sim \frac{1}{r}\exp(-r/\lambda_g)$, weakening over distances larger than the Compton wavelength of the graviton $r\gtrsim \lambda_g = \hbar/(m_gc^2)$. 
Hence, if the mass of the graviton is $m_g\sim H_0$ then gravity weakens at late times and on cosmological scales, causing an acceleration of the cosmic expansion relative to the GR prediction. 
The program to apply massive gravity as a dark energy model has hit important barriers, as flat FLRW solutions do not exist in this theory \cite{DAmico:2011eto}. 
Accelerating solutions without a cosmological constant (CC) do exist with open spatial hypersurfaces \cite{Gumrukcuoglu:2011ew}, but they are unstable \cite{DeFelice:2013awa}.
Proposed solutions include the graviton mass being generated by the vacuum expectation value of a scalar \cite{DAmico:2011eto} or deformations of the theory in which the BD ghost is introduced, which provides dynamical accelerating, but meta-stable solutions \cite{Konnig:2016idp}. 
Alternatively, one could promote the coefficients of the potential to be functions of the Stueckelberg fields~\cite{deRham:2014gla}.
Other ways to make massive gravity dynamical include the addition of a new field, such as a scalar field, e.g. quasi-dilaton \cite{DAmico:2012hia}, or one (or several) dynamical tensors in bigravity (and multigravity).

\paragraph{Bigravity and Multigravity}
In order to write a mass term for the metric, dRGT incorporates an additional, non-dynamical tensor, akin to the occurrence of $\eta_{\mu\nu}$ in eq. (\ref{eq:L_fierz_pauli}). 
Massive gravity can be extended by including a kinetic term to the auxiliary metric, which becomes fully dynamical. This leads to the theory of bigravity (or bimetric gravity) \cite{Hassan:2011zd}, which contains {two} spin-2 particles: one massive and one massless. The same procedure can be extended to more than two interacting metrics, leading to multigravity theories \cite{Hinterbichler:2012cn}. 
In these constructions there is always one massless excitation of the metric (a combination of the different tensor fields), with all other excitations being massive.

Bigravity solves the problem of cosmological evolution, at least at the background level. Flat FLRW solutions do exist, and many viable expansion histories have been found to be compatible with data \cite{Akrami:2012vf} and satisfying the Higuchi stability bound \cite{Fasiello:2013woa}. 
However, it was later found that these models had instabilities that affected the growth of linear perturbations \cite{Comelli:2014bqa}, which were found to be quite generic across different branches of solutions \cite{Konnig:2015lfa}. 
In some cases the instabilities affect only scales sufficiently small for non-linear effects to be important (i.e. the Vainshtein mechanism) which might render the theory stable \cite{Mortsell:2015exa}.
Another solution is to choose the parameters of the theory so instabilities occur at early times, when characteristic energies are high and bigravity is not a valid effective field theory. 
This happens by choosing a large hierarchy between the two Planck masses: the so-obtained theory is practically indistinguishable from GR plus a (technically natural) CC~\cite{Akrami:2015qga}.

\subsection{Descriptions of cosmological gravity}

\begin{figure*}[ht!]
\resizebox{.95\textwidth}{!}{
 \includegraphics[width=\textwidth]{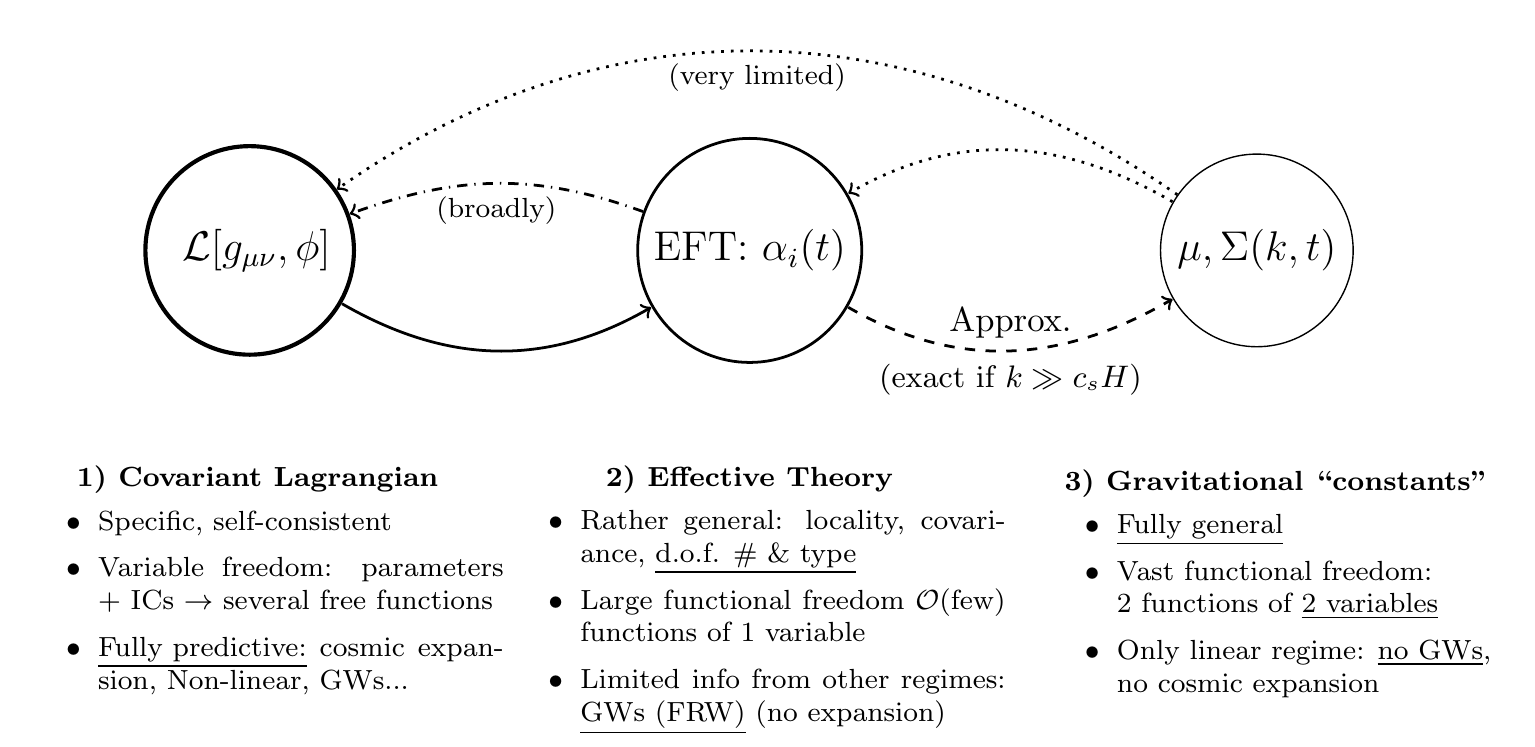}
}
\caption{Effective descriptions of cosmological gravity, their relations and main advantages/shortcomings.
Theories of gravity based on a gravitational Lagrangian are described in Sec. \ref{sec:covariant_theories}. 
The effective theory approach is described in \ref{sec:eft_de} and the Gravitational ''constants`` in section \ref{sec:g_effective}.
}\label{fig:effective_gravity}
\end{figure*}

The immense variety of alternative theories has motivated the search for effective descriptions able to capture the phenomenology of generic dark energy models. 
The covariant actions approach reviewed in Sec. \ref{sec:covariant_theories} offers several advantages, including 
1) full predictivity, as (classical) solutions can be found from microscopic scales, to strong gravity and all the way to cosmology,
2)~self-consistency, as different regimes can be computed for the same theory, leading to tighter constraints when the data is combined.
For instance, following this approach, we discuss the cosmology of covariant Galileons in Sec. \ref{sec:Hubbletension}.
Nonetheless, a great downside of this approach is that the predictions for every model/theory have to be obtained from scratch, which makes the exploration of the theory space a daunting task.

An alternative route is to constrain deviations from GR, without reference to any fundamental theory. 
The tradeoff is to keep the theory of gravity as general as possible at the expense of dealing with a very simple space-time. 
The simplest situation is where the background space-time is flat and maximally symmetric (Minkowski), a setup useful to model gravity in the Solar System. 
In this simple case one can define a series of quantities, known as Parameterized Post-Newtonian (PPN) coefficients, that describe general modifications of gravity over Minkowski space (see Ref. \cite{Will:2014kxa} for details, including constraints and additional assumptions). These PPN parameters that can be constrained by experiments (such as the deflection of light by massive bodies) and computed for any theory, and thus provide a very efficient phenomenological dictionary.

In cosmology we are interested in describing gravity over a slightly less symmetric background: a spatially homogeneous and isotropic, but time evolving, Friedmann-Lemaitre-Robertson-Walker (FLRW) metric:
\begin{equation}\label{eq:newt_gauge}
 ds^2 = -(1+2\Psi)dt^2 + a^2(t)\left\{(1-2\Phi)\delta_{ij} + h_{ij}\right\}dx^i dx^j \,,
\end{equation}
where  metric perturbations are in Newtonian gauge with the sign conventions of Ma \& Bertschinger \cite{Ma:1995ey}. The tensor perturbation is symmetric, transverse and traceless ($\partial^i h_{ij},\delta^{ij}h_{ij} = 0$) and we have ignored vector perturbations.
The time-evolution of the cosmological background makes an extension of  PPN approach to cosmology a difficult task, as instead of constant coefficients one needs to deal with functions of time due to the evolution of the universe.

The most important example of an effective description in cosmology is the parameterization of the cosmological background, often done in terms of the equation of state $w\equiv p/\rho$ \cite{Chevallier:2000qy,Linder:2002et}. Instead of computing the modifications to the Friedmann equations and the pressure and energy density contributed by the additional fields, a general approach to cosmological expansion is to specify $w(z)$ so that
\begin{eqnarray}
H^2 &=& \frac{8\pi G}{3}(\rho_M + \rho_{\rm DE})\,, \label{eq:Friedmann} \\
\rho_{\rm DE} & = & \exp\left( -3\int d\log(a) (1+w) \right)\,. \label{eq:w_de}
\end{eqnarray}
This is sufficient to describe any cosmological expansion history and in any theory (as long as matter is minimally coupled) just by using the Friedmann equation (\ref{eq:Friedmann}) as a definition for $\rho_{\rm DE}$.

Describing the perturbations requires more functional freedom. Here we will review two common procedures, namely the effective theory of dark energy and the modified gravitational ``constants''. The different approaches (including the covariant theory approach), their features and connections are outlined in Fig. \ref{fig:effective_gravity}. Consistency checks between the background and perturbations can also be used to test the underlying gravity theory \cite{Ruiz:2014hma,Bernal:2015zom}.

\subsubsection{Effective theory of Dark Energy}\label{sec:eft_de}

The effective (field) theory of dark energy (EFT-DE) \cite{Gubitosi:2012hu,Bloomfield:2012ff,Gleyzes:2013ooa} can be used to systematically describe general theories of gravity over a cosmological background (see Ref. \cite{Gleyzes:2014rba} for a review). 
The original formulation applies to theories with a scalar field $\phi$ and uses the unitary ``gauge'': 
a redefinition of the time coordinate as the constant $\phi$ hypersurfaces (this is always possible if $\phi_{,\mu}$ is time-like and non-degenerate, as in perturbed cosmological backgrounds, but not in general). 
One then constructs all the operators compatible with the symmetries of the background (recalling that the time translation invariance is broken by the cosmological evolution).

A very convenient basis for the EFT functions was proposed by Bellini \& Sawicki \cite{Bellini:2014fua}, when restricted to Horndeski's theory. 
In their approach the EFT functions are defined by the kinetic term of the propagating degrees of freedom in the equations of motion. The dynamical equation for tensor perturbations
\begin{equation}\label{eq:gw_horndeski_cosmo}
  \ddot h_{ij} + (3+\alpha_M) \dot h_{ij} + (1+\alpha_T)\frac{k^2}{a^2} h_{ij} = 0\,,
\end{equation}
introduces two dimensionless functions
\begin{itemize}
 \item \textit{tensor speed excess} $\alpha_T$ describes the modification in the GW propagation speed $c_{g}^2 = (1+\alpha_T)$. This modification is frequency independent (see Sec. \ref{sec:GWspeed}).
 \item  \textit{Planck-mass run rate} $\alpha_M$ enters as a friction term. It is related to the \textit{cosmological strength of gravity} $M_*^2$ (the kinetic term of tensor perturbations) by $\alpha_M = \frac{d\log(M_*^2)}{d\log a}$ (see Sec. \ref{sec:GWdamping}).
\end{itemize}
 The equations in the scalar sector (eqs. (3.20), (3.21) of \cite{Bellini:2014fua}) can be used to define the remaining functions. If we look only at the second time derivatives (that is, the kinetic terms)
\begin{eqnarray}\label{eq:scalar_horndeski_cosmo}
 2\ddot \Phi - \alpha_B H \delta\ddot{\phi}/\dot\phi\, + &\cdots& = 0 \,, \quad\,  \text{(ii-trace)} \\
 \alpha_K \delta\ddot{\phi}/\dot\phi + 3\alpha_B \ddot\Phi/H\, + &\cdots& = 0  \,, \quad \text{($\phi$ scalar)}
\end{eqnarray}
 (note the ellipsis denote terms without second time derivatives) one can define
 \begin{itemize}
  \item  \textit{braiding}, or \textit{kinetic gravity brading} $\alpha_B$ quantifies mixing between the second derivatives of the metric in the field equation (and vice versa). This is a generic property of modified gravity \cite{Deffayet:2010qz,Bettoni:2015wta}.
  \item \textit{kineticity} $\alpha_K$ modulates the ``stiffness'' of the scalar field (how hard it is to excite perturbations in $\phi$). The kineticity is intimately related to the propagation speed of scalar perturbations, which satisfies $c_s^2 \propto \alpha_K^{-1}$: higher kineticity values lead to slower scalar waves and vice versa.
 \end{itemize}
These functions can be computed from the Lagrangian functions in (\ref{eq:L_horndeski}), and for a given theory will depend on the value of the scalar field and its time derivative.
Constraints on the $\alpha$-functions can also be used to reconstruct the terms in a fundamental theory, as shown in Tab. \ref{tab:horndeski_eft}. Systematic reconstructions of the Lagrangian from the $\alpha$ functions have been also explored \cite{Kennedy:2017sof,Kennedy:2018gtx}.

\begin{table}[t!]
\begin{tabular}{c | c c c c c | c c c }
 & \multicolumn{5}{c}{Horndeski} & \multicolumn{3}{| c}{DHOST} \\
 & $G_{2,\phi}$ & $G_{2,X}$ & $G_{3,X}$ & $G_{4,\phi}$ & $G_{4,X}\cdots$ & GLPV & ${\cal C}_1$ & ${\cal C}_2$ \\ 
 \hline
 $1+w$ & $\checkmark$ & $\checkmark$ & $\checkmark$ & $\checkmark$ & $\checkmark$ & $\checkmark$ & $\checkmark$ & $\checkmark$  \\ \hline
 $\alpha_K$ & $-$ & $\checkmark$ & $\checkmark$ & $\checkmark$ & $\checkmark$ & $\checkmark$ & $\checkmark$ & $\checkmark$ \\
 $\alpha_B$ & $-$ & $-$ & $\checkmark$ & $\checkmark$ & $\checkmark$ & $\checkmark$ & $\checkmark$ & $\checkmark$ \\
 $\alpha_M$ & $-$ & $-$ & $-$ & $\checkmark$ & $\checkmark$  & $\checkmark$ & $\checkmark$ & $\checkmark$ \\
 $\alpha_T$  & $-$ & $-$ & $-$ & $-$ & $\checkmark$ & $\checkmark$ & $\checkmark$ & $\checkmark$ \\ \hline 
 $\alpha_H$  & $-$ & $-$ & $-$ & $-$ & $-$  & $\checkmark$ & $\checkmark$ & $\checkmark$ \\  
 $\beta_1$  & $-$ & $-$ & $-$ & $-$ & $-$  & $-$ & $\checkmark$ & $\bullet$ \\
 $\alpha_L$  & $-$ & $-$ & $-$ & $-$ & $-$  & $-$ & $-$ & $\checkmark$ \\ \hline 
\end{tabular}
\\ {\footnotesize $-$ zero, $\checkmark$ non-zero (arbitrary), $\bullet$ non-zero (constrained)}
\caption{EFT functions in scalar-tensor theories: a hyerarchy exists by which more complex theories of gravity (left to right) produce a larger set of effects (more non-zero functions). 
For the DHOST theories there are two classes of degeneracy conditions: ${\cal C}_1$ and  ${\cal C}_2$.
Some non-trivial special cases are known to exist: $f(R)$ and $f(G)$ theories have $\alpha_K=0$, while first generation theories (\ref{eq:lag_brans_dicke}) including $f(R)$, Brans-Dicke satisfy $\alpha_B+\alpha_M=0$ \cite{Bellini:2014fua} and 2 beyond Horndeski combinations produce $\alpha_T=0$ \cite{Creminelli:2017sry,Ezquiaga:2017ekz} (see Sec. \ref{sec:cg_consequences}).
\label{tab:horndeski_eft}
}
\end{table}

Increasingly complex theories of gravity lead to a larger number of EFT functions. 
In beyond Horndeski theories of the GLPV type, e.g. (\ref{eq:glpv4},\ref{eq:glpv5}), a new function $\alpha_H$ is introduced \cite{Gleyzes:2014qga}
which phenomenologically produces a weakening of gravity on small but linear cosmological scales \cite{DAmico:2016ntq}.
In DHOST theories including (\ref{eq:kin_conf_theory}) the situation is more involved, as the new EFT functions ($\alpha_L,\beta_1,\beta_2,\beta_3$) need to be related to each other and $\alpha_T,\alpha_H$ by the degeneracy conditions that prevent the introduction of additional degrees of freedom \cite{Langlois:2017mxy}. This leads to two classes of theories with one free function, which is either $\alpha_L$ or one among $\beta_i$.
New EFT functions appear beyond scalar-tensor theories, as has been explicitly derived for vector-tensor \cite{Lagos:2016wyv} and bimetric \cite{Lagos:2016gep} theories (including bimetric gravity), 
with a unified treatment of theories with different degrees of freedom \cite{Lagos:2017hdr}.

Different versions of the linear EFT-DE approach has been implemented in numerical codes able to obtain predictions based on linear perturbation theory. 
Publicly available implementations exist in EFTCAMB \cite{Hu:2013twa}, \texttt{hi\_class} \cite{Zumalacarregui:2016pph} and COOP \cite{Huang:2015srv}, with the first two based on the CAMB and CLASS Boltzmann codes \cite{Lewis:1999bs,Blas:2011rf}.
In addition, the CLASS-Gal code (integrated into CLASS) can be used to compute relativistic corrections to cosmological observables \cite{DiDio:2013bqa}.
These and other codes have been tested against a large class of models at a level of precision sufficient for current and next-generation cosmological experiments \cite{Bellini:2017avd}.

The EFT framework has been tested using linear observables. Horndeski theories were tested against current experiments, leading to $\mathcal{O}(0.1-1)$ constraints on the $\alpha$-functions varying over $\alpha_B,\alpha_M,\alpha_T$ \cite{Bellini:2015xja}, with $\alpha_M=-\alpha_B$ \cite{Ade:2015rim} and setting $\alpha_T=0$ to reflect the strong bounds on the GW speed \cite{Kreisch:2017uet} ($\alpha_K$ is very weakly constrained by current data). 
Future experiments have great potential to improve on these bounds, and are expected to improve the sensitivity to $\mathcal{O}(0.01-0.1)$ \cite{Gleyzes:2015rua,Alonso:2016suf,Lorenz:2017iez,Mancini:2018qtb,Reischke:2018ooh}.
EFT-based modifications of gravity might be observable through relativistic effects on ultra-large scales \cite{Renk:2016olm,Lorenz:2017iez} (see also the discussion in Sec. \ref{sec:g_effective}): these techniques might improve significantly our ability to constrain $\alpha_K$, although it will remain the hardest to measure \cite{Alonso:2016suf}.
Those works used simple functional dependence of the EFT functions. It has been nonetheless shown that simple parameterizations are indistinguishable from more complex models in most cases, even for next-generation cosmology experiments \cite{Gleyzes:2017kpi}. 

The EFT approach has been generalized beyond linear perturbations for Horndeski theories. 
Including non-linear cosmological perturbations in general introduces new functions at every order in perturbation theory (e.g. to compute the bispectrum \cite{Bellini:2015wfa}). 
However, a restriction to cubic and quartic operators (in the unitary gauge) leads to only 3 new operators on quasi-static scales \cite{Cusin:2017mzw}. 
Some applications of non-linear EFT-DE include corrections to the power spectrum (e.g. \cite{Cusin:2017wjg}), the use of higher-order correlations as a test of gravity, such as the bispectrum of matter \cite{Bellini:2015wfa}, galaxies \cite{Yamauchi:2017ibz} and CMB lensing \cite{Namikawa:2018erh} or the the non-linear shift of the BAO scale~\cite{Bellini:2015oua}.

\subsubsection{Modified Gravitational ``constants''} \label{sec:g_effective}

A very commonly used approach employs general modifications of the equations relating the gravitational potentials to the matter density contrast 
\begin{eqnarray} \label{eq:mu_eff}
&& \nabla^2\Psi = 4\pi G a^2 {\mu(t,k)}\rho \delta \,, 
\\ \label{eq:sigma_eff}
&&  \nabla^2(\Phi+\Psi) = 8\pi G a^2 { \Sigma(t,k)} \rho\delta \,
\end{eqnarray}
(note that different conventions exist in the literature).
Here $\delta$ is the density contrast in the Newtonian gauge (\ref{eq:newt_gauge}) and the functions $\mu,\Sigma$ parameterize the evolution of the gravitational potentials as a function of time $a$ and scale $k$.
The functions $\mu,\Sigma$ are often referred to as $G_{\rm matter}$, $G_{\rm light}$ because gradients of $\Psi$ determines the force felt by non-relativistic particles and those of $\Psi+\Phi$ the geodesics of massless particles (and thus the lensing potential).
The ratio of the gravitational potentials, 
\begin{equation}
\label{eq:eta_eff}
\eta \equiv \frac{\Phi}{\Psi} = \frac{2\Sigma}{\mu}-1\,,
\end{equation}
is of particular interest, since GR predicts that it is exactly one in the absence of radiation and any sizable deviation could be an indication of modified gravity.
 
This approach has numerous advantages as a test of gravity against data. It is completely theory agnostic, not requiring any information on the ingredients or laws of the theories being tested. 
Most importantly, it is completely general for universally coupled theories: given any solution $\Delta,\Psi,\Phi(a,k)$ it is possible to obtain $\mu,\Sigma$ through (\ref{eq:mu_eff},\ref{eq:sigma_eff}). 
In this sense, any finding of $\mu,\Sigma\neq 1$ might point towards deviations from GR and warrant further investigation.

The main shortcoming of this approach is its great generality: any practical attempt to implement (\ref{eq:mu_eff},\ref{eq:sigma_eff}) requires a discretization of the functional space, introducing $2\cdot N_k \cdot N_z$ free parameters for a homogeneous binning.
In contrast, the EFT approach for Horndeski theories (\ref{eq:gw_horndeski_cosmo},\ref{eq:scalar_horndeski_cosmo}) requires only $4\cdot N_z$ parameters, making it a more economic parameterization for all but the simplest scale-dependencies ($N_k = 1,2$). 
Capturing the full scale dependence of $\mu,\Sigma$ requires either a large parameter space or assumptions about the $k$-dependence.

A common practice to overcome this limitation is to choose a functional form for $\mu,\Sigma$ as a function of scale.
For Horndeski theories the functional form is a ratio of quadratic polynomials in $k$ \cite{DeFelice:2011hq,Amendola:2012ky}
\begin{equation}\label{eq:g_eff_horndeski}
 \mu = h_1\frac{1+ h_5 k^2}{1+ h_3 k^2} \,,\quad \eta = h_2\frac{1+h_4 k^2}{1+h_5 k^2}\,,
\end{equation}
for functions $h_i$ that depend on redshift through the theory (\ref{eq:L_horndeski}) and the scalar field evolution. The mapping is exact on small scales in which the field dynamics can be neglected, below scalar sound horizon \cite{Sawicki:2015zya}.
A $k$-dependence as the ratio of polynomials is generic in local theories at quasi-static scales \cite{Silvestri:2013ne}, with higher order polynomials possible in Lorentz-violating \cite{Baker:2014zva}, multi-field \cite{Vardanyan:2015oha} theories. 
Studies with current data have tested rather simple parameterizations of $\mu,\Sigma$: for instance the Planck survey tested the case of $k$-independent $\mu,\eta$ in addition to the theory-motivated (\ref{eq:g_eff_horndeski}) \cite{Ade:2015rim}.
Future surveys will improve the resolution on the scale-dependence: 3 $k$-bins are the minimum to constraint all the parameters in eq. (\ref{eq:g_eff_horndeski}), with  6 bins in $z$ \cite{Amendola:2013qna,Taddei:2016iku}. 
A limited handle on scale-dependence on ultra-large scales might be achievable \cite{Baker:2015bva,Villa:2017yfg} (see also \cite{Raccanelli:2013dza,Lombriser:2013aj,Bonvin:2018ckp} for related parameterizations).

Another main shortcoming of the completely general approach is that there is no information from other regimes. The major setback with respect to EFT is the lack of information from gravitational wave observables, while in EFT the tensor and scalar sectors are modified accordingly i.e. GW data restrict the modifications available to scalar perturbations, for instance, theories with $\eta\neq 1$ require either $\alpha_M$ or $\alpha_T$ to be non-zero \cite{Saltas:2014dha}.
Attempts to explore the connections between  $\mu,\Sigma$ and the EFT approach in Horndeski-like theories have used very general parameterizations:
connecting theoretical viability conditions of the theory with the behavior of $\mu,\eta$ \cite{Perenon:2015sla}, including the case with $\alpha_T=0$ to address the impact of the GW speed measurement \cite{Peirone:2017ywi}.
General properties of Horndeski theories could be inferred from detailed measurements of $\mu,\Sigma$ \cite{Pogosian:2016pwr}.
Similarly to the EFT approach, the background evolution is unknown and the equation of state (\ref{eq:w_de}) is in principle arbitrary. However, theoretical priors on $w(z)$ can be obtained for broad classes of Lagrangians (e.g. quintessence \cite{Marsh:2014xoa}) or from stability conditions in general realizations of the EFT functions \cite{Raveri:2017qvt}.
 
 \section{Basics of Gravitational Waves}
 \label{sec:BasicsGW}
 
Gravity is a universal, long-range force. This, in field theory language, implies that it must be described by a metric field $\gmn$ in order to manifestly preserve locality and Lorentz invariance. At low energies, the leading derivative interactions are second order. Therefore, gravity theories generically predict the existence of propagating perturbations or, in other words, the existence of GWs. One can define a metric perturbation $\hmn$ as a small difference between the metric field $\gmn$ and the background metric $\gbmn$
 \be
 \hmn=\gmn-\gbmn\,,
 \ee
 where $\vert\hmn\vert\ll1$. However, in curved space it is non-trivial to distinguish the perturbation from the background unless the latter posses some degree of symmetry, e.g. flat space or FLRW. A way out is to define GWs via geometric optics \cite{misner1973gravitation}. In this context, the key element to distinguish the GW from the background is the size of the fluctuations $\lGW$ with respect to the typical size of the background variation $L_{_\text{B}}$. One could associate the typical variation scale in the background with the minimum value of the components of the background Riemann tensor
 \be
 L_{_\text{B}}\sim\vert R^{_\text{B}}_{\alpha\beta\gamma\rho}\vert^{-1/2}\,.
 \ee
 For astrophysical sources, we will see later that the wavelength of the GW $\lGW$ is orders of magnitude smaller than the typical variations of $L_{_\text{B}}$ for cosmological setups. The fact that $\lGW\ll L_{_\text{B}}$ implies that there is a clear hierarchy between background and perturbations, allowing to solve the problem using an adiabatic (or WKB) expansion.
 
In the following, we describe the basics of GWs. We begin by introducing GWs in GR. Then, we explore the propagation in cosmological backgrounds. Subsequently, we describe how this picture is changed beyond GR. Finally, we discuss the status of present and future GW detectors. We recommend the reader Ref. \cite{misner1973gravitation,maggiore2008gravitational,Maggiore:2018sht,Flanagan:2005yc,Carroll:2004st} for more details.
 
 \subsection{GWs in General Relativity}
 
 General Relativity is a universal, infinite-range force. As we have seen in the previous section, this implies that it is described by a massless, spin-2 field. 
The dynamics is described by Einstein's equations (\ref{eq:EinsteinEquation}). Importantly, not all the components of the Einstein tensor $G_{\mu\nu}$ contain second order time derivatives of the metric $g_{\mu\nu}$. This implies that not all of the 10 components of $g_{\mu\nu}$ will propagate. In particular, the $G_{0\mu}$ equations act as 4 constraint equations. This, together with the 4 unphysical modes reduced by the gauge choice, leaves only 2 propagating degrees of freedom. This is precisely what one would expect for a massless spin-2 particle.
 
In order to study GWs, the next step is to study the linearized Einstein's equations. To diagonalize the equations for the tensor perturbations, one has to introduce the trace-reversed perturbation 
 \be
 \thmn=\hmn-\frac{1}{2}h\gbmn\,,
 \ee
 whose name comes from the fact that $\bar{h}=-h$ where $h=g_{_\text{B}}^{\mu\nu}\hmn$ and $\bar{h}=g_{_\text{B}}^{\mu\nu}\thmn$ are the traces of $\hmn$ and $\thmn$ respectively. Fixing the Lorenz gauge for this new variable $\nabla^{\mu}\thmn=0$, the linearized Einstein equations in curved space-time read
 \be
 \label{eq:PropCurvedSpave}
 \begin{split}
 \Box\thmn&+2R^{_\text{B}}_{\mu\alpha\nu\beta}\bar{h}^{\alpha\beta}=\\
 -16\pi G&\delta T_{\mu\nu}+2R^{_\text{B}~\alpha}_{~(\mu}\bar{h}_{\nu)\alpha}-R^{_\text{B}}h_{\mu\nu}+\gbmn R^{\alpha\beta}_{_\text{B}}\bar{h}_{\alpha\beta}\,,
 \end{split}
 \ee
where covariant derivatives are built with the background metric $\gbmn$. Here, we have introduced the perturbed energy-momentum tensor $\delta T_{\mu\nu}$ as the difference of the total energy momentum tensor $T_{\mu\nu}$ with respect to the background solution $8\pi GT^{_\text{B}}_{\mu\nu}=R^{_\text{B}}_{\mu\nu}-\frac{1}{2}g^{_\text{B}}_{\mu\nu}R^{_\text{B}}$. One should note that, in vacuum, all the Ricci tensors vanish in the second line. Moreover, for short-wave GWs $\lGW\ll L_{_\text{B}}$, the Riemann tensor in the first line has a subdominant contribution.
 
To deal with the two GW polarizations, it is convenient to work in the transverse-traceless (TT) gauge, which is defined by
 \be
 h_{0\mu}=0\,,\quad \nabla^{j}h_{ij}=h^{i}_{~i}=0\,.
 \ee
 Note that in the TT gauge, the trace-reversed perturbation $\thmn$ is equal to the original perturbation $\hmn$. If the GW is propagating in the $z$-direction, the spatial components become 
 \be
 \label{eq:GRpolarizations}
 h_{ij}=\begin{pmatrix} h_{+} & h_\times & 0 \\ h_\times & -h_+ & 0 \\ 0 & 0 & 0 \end{pmatrix}\,,
 \ee
with $h_+$ and $h_\times$ being the two polarizations of GR. 
 \subsubsection{Generation}
 
A first question to address is how GWs are produced. Let us consider a GW source in vacuum within the short-wave approximation. Then, the general propagation equation (\ref{eq:PropCurvedSpave}) reduces to 
 \be
 \Box\thmn=-16\pi G\,\delta T_{\mu\nu}\,.
 \ee
 This wave equation can be solved in analogy to electromagnetism using a Green's function. In terms of the retarded time $t_r=t-\vert\vec{x}-\vec{y}\vert$, the solution is 
 \be
 \thmn(t,\vec{x})=2G\int d^3\vec{y}\, \frac{\delta T_{\mu\nu}(t_r,\vec{y})}{\vert\vec{x}-\vec{y}\vert}\,.
 \ee 
 For an isolated, far away, non-relativistic source, this solution can be simplified. In fact, one can make a multipole expansion. The zeroth moment corresponds to the mass-energy of the source $M=\int T^{00}(t,\vec{y})d^3\vec{y}$. However, conservation of energy for an isolated source tells us that $M$ cannot vary in time. Next, the mass dipole moment $M_{i}(t)=\int y_i\,T^{00}(t,\vec{y})d^3\vec{y}$ is associated to the motion of the center of mass. Nevertheless, its time derivative is the momentum of the source that also has to be conserved\footnote{Similar arguments apply for the spin angular momentum in case the source exhibit some internal motion.}. Consequently, the leading contribution is the mass quadrupole moment $M_{ij}(t)=\int y_iy_j\,T^{00}(t,\vec{y})d^3\vec{y}$, which generates GWs through its second time derivatives
 \be
 \bar{h}_{ij}(t,\vec{x})=\frac{2G}{r}\frac{d^2 M_{ij}}{dt^2}(t_r)\,.
 \ee 
 
\begin{figure}[t!]
\centering 
 \includegraphics[width=0.99\columnwidth]{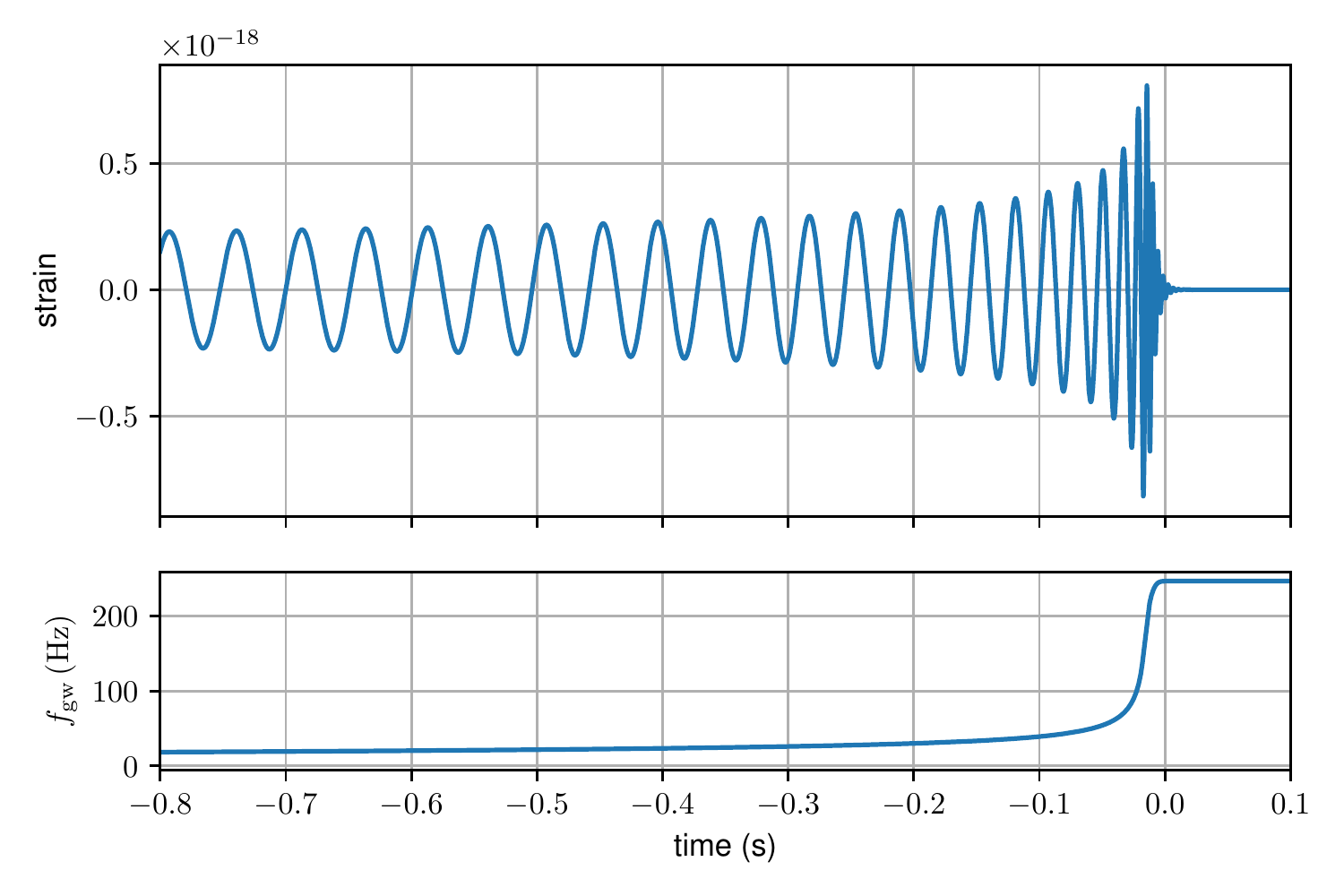}
 \caption{Typical GW signal of a compact binary coalescence. The GW strain (above) and the GW frequency (below) are plotted as function of the time before merging. This waveform is a template of the first event detected GW150914 \cite{Vallisneri:2014vxa}.}
 \label{fig:GWtemplate}
\end{figure}

 For a binary system of masses $m_1$ and $m_2$, the quadrupole radiation is 
 \be
 \label{eq:WaveForm}
 h_{+,\times}=\frac{\mathcal{M}_c^{5/3}f^{2/3}}{r}F_{+,\times}(\mathrm{angles})\cos\Phi(t)\,,
 \ee
 where $F$ is a function of the orientation of the binary that depends on the polarization $+$ or $\times$ (recall (\ref{eq:GRpolarizations})), $\Phi(t)$ is the phase and we have introduced the chirp mass
 \be
 \mathcal{M}_c=\frac{(m_1m_2)^{3/5}}{(m_1+m_2)^{1/5}}\,.
 \ee
 As the masses orbit one around the other, they will lose energy with the emission of GWs. They will begin getting closer and orbiting faster until they eventually merge. Thus, the frequency of GWs will increase with a characteristic chirp signal following
 \be
 \label{eq:GWFrequency}
 \dot{f}_{\mathrm{gw}}=\frac{96}{5}\pi^{8/3}\left(\frac{G\mathcal{M}_c}{c^3}\right)^{5/3}f_{\mathrm{gw}}^{11/3}\,.
 \ee
 Note that to consider the energy loss due to GWs emission one has to go to second order in perturbation theory. An example of the typical GW strain and frequency of a compact binary coalescence is presented in Fig. \ref{fig:GWtemplate}.
 
 Typical binary compact objects emitting detectable GWs are binary neutron stars (BNS) and binary black-holes (BBH). The order of magnitude of the frequency of the GWs of these systems is
 \be
 f_{\mathrm{gw}}\sim \frac{1}{4\pi}\lp\frac{3GM}{R^{3}}\rp^{1/2}\sim 1\text{kHz}\lp\frac{10M_\odot}{M}\rp\,,
 \ee
where $M_\odot$ is equal to one solar mass. This implies that higher masses lead to lower frequencies. In terms of the wavelength one finds
 \be
 \lGW\sim200\text{km}\lp\frac{M}{M_\odot}\rp\,.
 \ee
 This allows us to compare the size of the wavelength with the typical size of the background curvature variation $L_{_\text{B}}$. For cosmology, the size of the curvature is related to the Hubble horizon $L^{\text{cosmo}}_{_\text{B}}\sim 10^{26}$m. For our galaxy one can estimate $L^{\text{gal}}_{_\text{B}}\sim10^{23}$m and for the Solar System $L^{\text{SolSys}}_{_\text{B}}\sim10^{16}$m.
As it can be observed, the geometric optics expansion is an excellent approximation due to the great hierarchy between $\lGW$ and $L_{_\text{B}}$. Only GWs passing near a very dense object such as a BH, $L^{\text{BH}}_{_\text{B}}\sim(M_{_\text{BH}}/M_\odot)$km, would break this short-wave approximation.
 
 The typical amplitude of a GW from a compact binary can be estimated using (\ref{eq:WaveForm}), leading to 
 \be
 h\sim10^{-21}\lp\frac{\mathcal{M}_c}{10\,M_\odot}\rp^{5/3}\lp\frac{f}{100\,\text{Hz}}\rp^{2/3}\lp\frac{100\,\text{Mpc}}{r}\rp\,.
 \ee
 Contrary to EM waves, GW detectors are directly sensitive to the amplitude of the wave, which falls like $1/r$ and not as the luminosity $1/r^2$. This means that even if the amplitudes are very small, GW detectors are more sensitive to distant sources.
 
 \subsubsection{Propagation}
 
 Once the GW is generated, it will propagate in vacuum following
 \be
 \label{eq:propGW}
 \Box\thmn+2R^{_\text{B}}_{\mu\alpha\nu\beta}\bar{h}^{\alpha\beta}=0\,.
 \ee
 A general solution of this wave equation can be written as the sum of plane waves
 \be
 \thmn(t,\vec{x})=\text{Re}\lb A_{\mu\nu}\cdot e^{ix_\alpha k^\alpha}\rb\,,
 \ee
 where $\text{Re}$ denotes the real part. By plugging this expression in the wave equation and expanding in powers of $k$, one finds at leading order that
 \be
 \label{eq:dispersionGR}
 k_\mu k^\mu=g_{_\text{B}}^{\mu\nu}k_\mu k_\nu=0\,.
 \ee
 Therefore, GWs propagate in null geodesics determined by the background metric. This means that the GW-cone is the same as the light-cone and both waves propagate at the same speed. Moreover, the wave is transverse to the propagation direction
 \be
 k^\mu A_{\mu\nu}=0\,,
 \ee
 similarly to electromagnetic waves.
 Finally, by defining the scalar amplitude $\A=\lp \frac{1}{2}A^{*}_{\mu\nu}A^{\mu\nu}\rp^{1/2}$ one realizes that
 \be
 \nabla_\alpha\lp k^\alpha \A\rp=0\,,
 \ee
 which can be interpreted as the conservation of gravitons. One should note that $R^{_\text{B}}_{\mu\alpha\nu\beta}$ in the wave equation only modifies the amplitude at second order. Consequently, at first order in geometric-optics, the wave equation $\Box\thmn=0$ can be rewritten as
 \be
 \Box R^{\text{gw}}_{\mu\alpha\nu\beta}=0\,.
 \ee
 This expression could be used as a gauge invariant, coordinate independent definition of the propagation of GWs in vacuum.
 
 \subsubsection{Detection}
  
\begin{figure*}[t!]
\centering 
\includegraphics[width=0.9\textwidth]{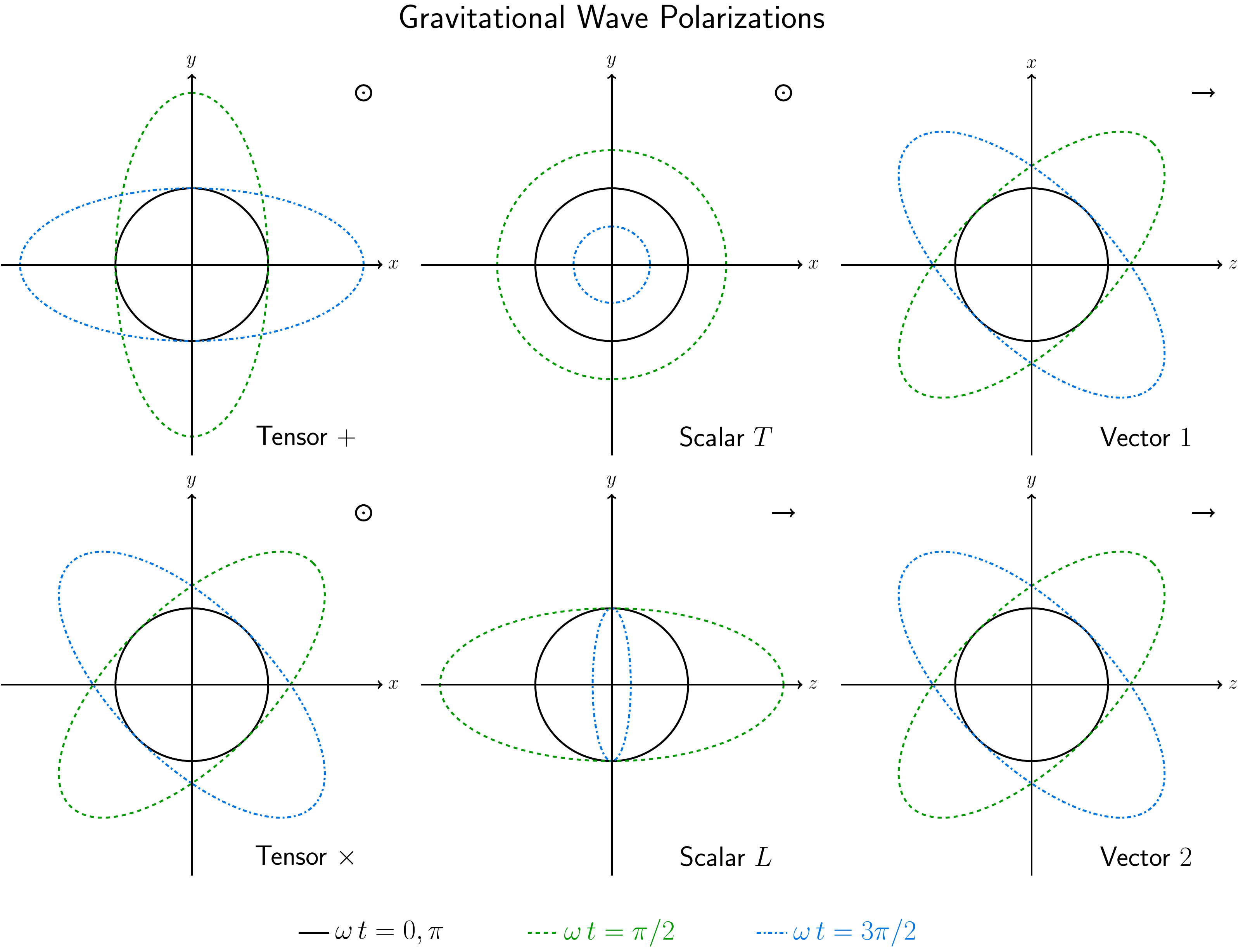}
\caption{Possible gravitational wave polarizations. A circle of test masses is distorted differently for each polarization propagating on the $z$-direction as a function of time ($\omega\,t=0,\pi/2,\pi,3\pi/2$). General Relativity only contains the two tensor polarizations $+$ and $\times$. Other gravity theories might contain also a transverse (breathing) scalar mode (Scalar $T$), a longitudinal scalar (Scalar $L$) and two vector modes (Vector $1\,,2$).}
  \label{fig:Polarizations}
 \end{figure*}

 To see the effect of a GW passing by, one has to study the deviation of nearby geodesics. Given two particles with four-velocity $U^\mu$ separated by $S^\mu$, their separation evolves as
 \be
 \frac{D^2S^\mu}{d\tau^2}\equiv U^\rho\nabla_\rho\lp U^\gamma\nabla_\gamma S^\mu\rp=R^\mu_{~\alpha\beta\nu}U^\alpha U^\beta S^\nu\,,
 \ee
 where $\tau$ is the proper time. At leading order, the four velocity is just the unit vector $U^\mu=(1,0,0,0)+\mathcal{O}(h)$, and we only have to compute the Riemann tensor in the TT gauge. The result is 
 \be
 \frac{\partial^2S^\mu}{\partial t^2}=\frac{1}{2}S^\nu\frac{\partial^2}{\partial t^2} h^{\mu}_{~\nu}\,,
 \ee
 where we have also used that to leading order the proper time $\tau$ and the coordinate time $t$ coincide. Accordingly, only the components of the separation vector $S^\mu$ transverse to the propagation vector will feel the effect of the GW. In these directions, the separation between the test particles will oscillate as the GW travels perpendicular to them. In Fig. \ref{fig:Polarizations}, we plot the effect of the different GW polarizations crossing a circle of test masses.
 
 GW detectors precisely rely on this principle that GWs can alter the separation between test masses. Modern detectors are interferometers. In brief, they are constituted by two perpendicular arms of the same length with two mirrors in free fall at their ends (acting as test particles). A laser beam is split in the two arms so that the beams reflect in each mirror and come back to the splitting point. In the absence of a GW, both laser beams returning will interfere destructively and no signal would arrive to the detector. However, if a GW crosses the interferometer, it will change the length of the arms differently. This means that the laser beams will take different times to travel the arms, arriving at the splitting point with different phases. Then, the destructive interference is lost and some signal gets to the detector. 
 
 Note that the typical distance variation $\delta L$ of two test masses separated by $L$ is approximately $\delta L\sim h\cdot L$. For compact binaries, we have seen that the strain amplitude is $h\sim10^{-21}$. Therefore, LIGO-type detector with arms of the order of kilometers have to measure distance variations
 \be
 \delta L\sim 10^{-18}\lp\frac{h}{10^{-21}}\rp\lp\frac{L}{\text{km}}\rp\text{m}\,,
 \ee
 a thousand times smaller than the nucleus of an atom. To achieve that, each arm has a resonant cavity in which the laser beams bounce back and forth about 300 times. This effectively makes ground-based interferometer arms to be $1200$km long (since the variation time of the GW is much longer than the travel time of the laser in the cavity). Accordingly, LIGO is sensitive to frequencies of $f_{_\text{LIGO}}\sim10^2$Hz. For the future space-based interferometer LISA, the working principle will be the same but with longer arms $L\sim10^{6}$km, being thus sensitive to much smaller frequencies, $f_{_\text{LISA}}\sim10^{-2}$Hz.
 
 \subsection{GWs in cosmology}
 \label{sec:GWcosmology}
 
At large scales, the universe is homogeneous and isotropic to very high accuracy. The background geometry is then described by a (flat) Friedmann-Lemaitre-Robertson-Walker (FLRW) metric
 \be
 ds^{2}=\gbmn dx^\mu dx^\nu=a^{2}(\eta)\lp-d\eta^2+d\vec{x}^2\rp\,,
 \ee
 where $a(\eta)$ is the scale factor and we are timing in conformal time $d\eta=dt/a(t)$. The propagation equation (\ref{eq:propGW}) becomes in Fourier space
 \be
 \label{eq:CosmoProp}
 h''_{ij}+2\mathcal{H} h'_{ij}+k^2h_{ij}=0\,,
 \ee
 where $\mathcal{H}=a'/a$ is the Hubble parameter and primes denote derivatives with respect to conformal time. This is nothing but a wave equation with a friction term due to the cosmic expansion. This Hubble friction will produce a redshift of the frequencies $f^\text{emit}=(1+z)f^\text{obs}$ and a rescaling of the GW amplitude $h\sim1/(a\cdot r)$. The previous formulae for a compact binary (\ref{eq:WaveForm}-\ref{eq:GWFrequency}) written in terms of the observed frequency $f^\text{obs}$ are thus valid if we replace the chirp mass $\mathcal{M}_c$ by the redshifted chirp mass 
 \be
 \mathcal{M}_z=(1+z)\mathcal{M}_c
 \ee
 and the physical distance $a\cdot r$ by the GW luminosity distance 
 \be
 \label{eq:luminositydistance}
 d_{L}^{\mathrm{gw}}=(1+z)\int_0^z\frac{c}{H(z)}dz\,,
 \ee
 where $c$ is the speed of light and $z$ the redshift. In this way, all the $(1+z)$ terms cancel each other. Note that there is an intrinsic degeneracy between the redshift and the Hubble parameter $H(z)$ in the GW luminosity distance. Therefore, the expansion history can only be obtained from the GW amplitude if the redshift is known. For near by sources $z\ll1$, the Hubble constant $H_0$ can be obtained
 \be
 \label{eq:luminositydistanceExpansion}
 d_{L}^{\text{gw}}=\frac{cz}{H_0}+\mathcal{O}(z^2)\,,
 \ee
 showing the power of GW astronomy to do cosmology. We will review this topic in more detail in Sec. \ref{sec:StandardSirens}.
 
Finally, let us mention that we have only focused on GWs from binary sources in the late universe. However, there could be other sources of GWs in the early universe leading to stochastic, cosmological backgrounds. For a nice review in the subject one can follow \cite{Caprini:2018mtu}. One may wonder if there could be an effect in the GW propagation when traveling through the cold dark matter. This question has been addressed recently and the answer is that the effect is too small \cite{Baym:2017xvh,Flauger:2017ged}.

 \subsection{GWs beyond GR}
 
 As we have emphasized at the beginning of this section, the existence of wave solutions for metric perturbations is generic for second order gravity theories. However, the behavior of these GWs can be very different depending on the gravity theory. The differences can arise either at the production or the propagation.
 
 \subsubsection{Additional polarizations}
  
During the generation of GWs, the main differences in theories beyond GR is that there could be other polarizations excited. We have seen that in GR only the 2 tensor polarizations propagate (recall (\ref{eq:GRpolarizations})). Nevertheless, modifications of gravity might introduce new degrees of freedom. For instance, in scalar-tensor theories there will be an additional scalar mode. Or in Massive Gravity, where there will be in addition 2 vector modes and a scalar one. For a GW propagating in the $z$-direction, one could decompose the amplitude $A_{ij}$ in the different polarizations 
 \be
 \label{eq:polarizations}
 A_{ij}=\begin{pmatrix} A_S+A_{+} & A_\times & A_{V1} \\ A_\times & A_S-A_+ & A_{V2} \\ A_{V1} & A_{V2} & A_L \end{pmatrix}\,,
 \ee
 where $A_+$ and $A_\times$ are the two tensor modes, $A_{V1,2}$ the two vector polarizations, $A_S$ the transverse (breathing) scalar and $A_L$ the longitudinal scalar mode. 
 One should note that these other types of polarizations will also leave an imprint in the detectors. Each polarization will have a different effect as we exemplify in Fig. \ref{fig:Polarizations}. In principle, with a set of 6 detectors one could distinguish all possible polarizations. 
 
Before continuing, it is important to remark that if a source is emitting additional polarizations, it will lose energy more rapidly. For a binary pulsar, if additional modes were emitted, the orbit would shrink faster due to the higher energy loss. For PSR B1913+16 (better known as Hulse-Taylor pulsar) \cite{Hulse:1974eb}, the orbit has been tracked for more than four decades now, showing an impressive agreement with GR \cite{Weisberg:2010zz}. Binary pulsars have been intensively used to constrain alternative theories of gravity, placing severe bound on dipolar radiation as reviewed in \cite{Stairs:2003eg,Wex:2014nva}. An example of this are Einstein-Aether propagating waves \cite{Jacobson:2004ts}, which have been constrained from pulsars due to dipolar GW emission \cite{Yagi:2013qpa,Yagi:2013ava}. Another would be the constraints on Brans-Dicke from a pulsar-white dwarf binary \cite{Freire:2012mg}.

Due to these constraints on the emission of additional polarizations, it is usually invoked a screening mechanism around the source to evade them. If this is the case, deviations of GR could only be measured in the propagation of GWs. We will discuss more about the emission of extra modes and screening mechanisms in Sec.~\ref{sec:CompactObjects}.
 
\subsubsection{Modified propagation}

 The propagation of GWs in gravity theories beyond GR can be very complicated. The additional fields might modify the background over which GWs propagate and their perturbations could even mix with the metric ones. For simplicity, we will restrict here to cosmological backgrounds. In that case, due to the symmetries of FLRW, tensor perturbations can only mix with other tensor perturbations. Possible deviations from the cosmological wave equation in GR (\ref{eq:CosmoProp}) can be parametrized by \cite{Nishizawa:2017nef}
 \be
 \label{eq:ModProp}
 h''_{ij}+(2+\nu)\mathcal{H} h'_{ij}+(c_g^2k^2+m^2a^2)h_{ij}=\Pi_{ij}\,,
 \ee
 where $\nu$ is an  additional friction term, $c_g$ accounts for an anomalous propagation speed, $m$ is an effective mass and $\Pi_{ij}$ is a source term originated by the additional fields. For instance, the scalar-tensor analogue of this equation is (\ref{eq:gw_horndeski_cosmo}). It is interesting that the modified GW propagation can also be understood in analogy with optics as GWs propagating in a diagravitational medium \cite{Cembranos:2018lcs}. 
 
 Focusing on the case without sources, $\Pi_{ij}=0$, the original GR wave-form $h_{_{\mathrm{GR}}}$, given by (\ref{eq:WaveForm}) for instance, will be modified by
 \be
 h_{_{\mathrm{GW}}}\sim h_{_{\mathrm{GR}}}\,\underbrace{e^{-\frac{1}{2}\int  \nu\mathcal{H}d\eta}}_\text{Affects amplitude}\,\underbrace{e^{ik\int(\alpha_{_T}+a^2m^2/k^2)^{1/2}d\eta}}_\text{Affects phase}\,,
 \ee
 where we have introduced $\aT=c_g^2-1$. Mainly, the additional friction will modify the amplitude, while the anomalous speed and the effective mass change the phase. The modified luminosity distance is then\footnote{See Appendix A of the first arXiv version of \cite{Ezquiaga:2017ekz} for a derivation.}
 \be \label{eq:MGluminosity}
 d_{L}^{_\text{MG}}=(1+z)\frac{c_g(z)}{c_g(0)}\text{exp}\lb\frac{1}{2}\int_0^z  \frac{\nu}{1+z'}dz'\rb\int_0^z\frac{c_g(z')}{H(z')}dz'\,.
 \ee
We will discuss how to test the GW phase in Sec. \ref{sec:GWspeed} and the damping of the strain in Sec. \ref{sec:GWdamping}.
 
\begin{figure*}[t!]
\centering 
\includegraphics[width=0.9\textwidth]{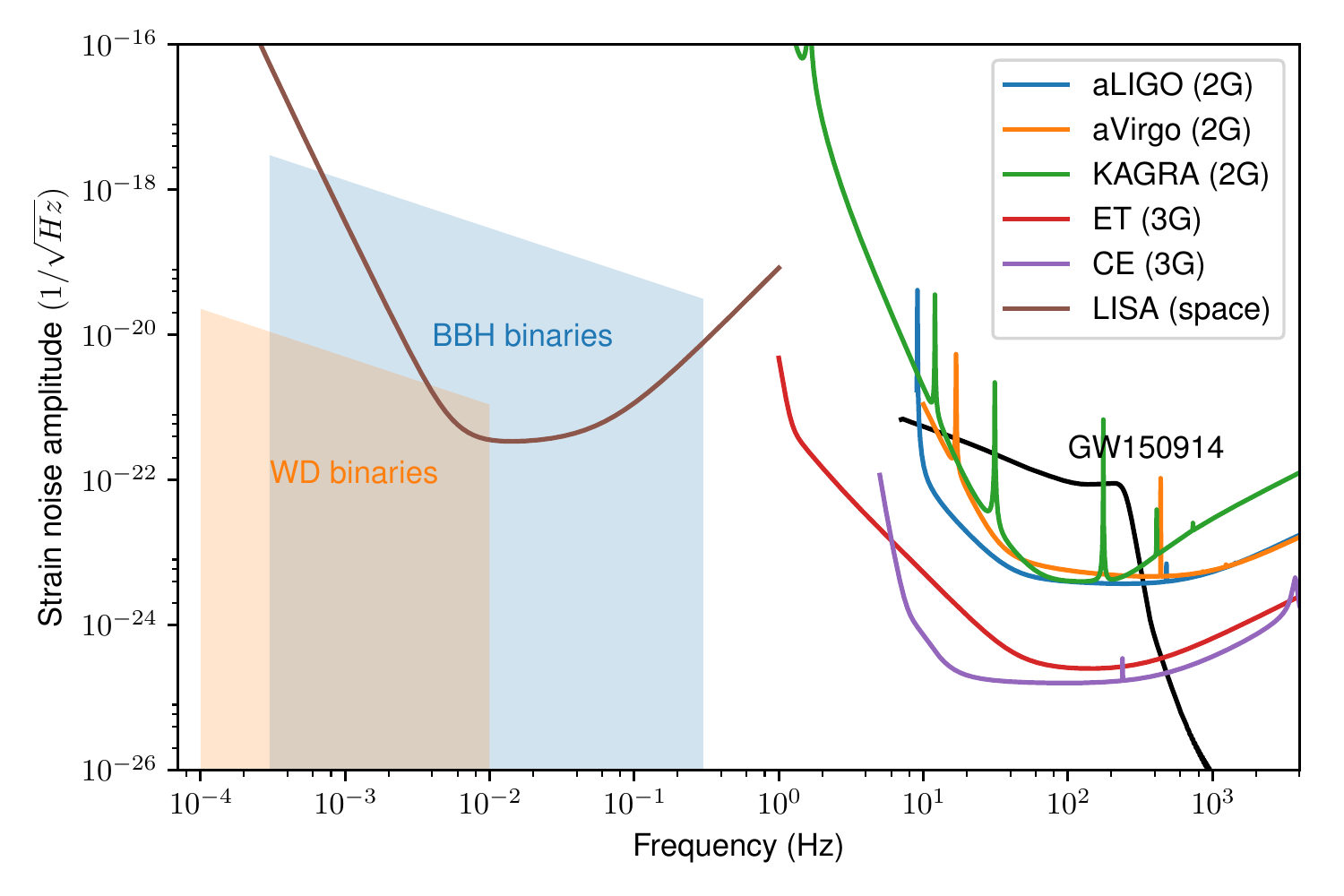}
\vspace{-10pt}
\caption{Strain sensitivity curves for different GW detectors. Second generation (2G) ground-based detectors are advanced LIGO (aLIGO), advanced Virgo (aVirgo) and KAGRA, with curves given at design sensitivity \cite{Aasi:2013wya}. Third generation (3G) detectors projected are Einstein Telescope (ET) \cite{Sathyaprakash:2012jk} and Cosmic Explorer (CE) \cite{Evans:2016mbw}. A space-based detector planned is LISA \cite{AmaroSeoane:2012km}. For illustration, we include the strain amplitude of GW150914 \cite{Vallisneri:2014vxa} and the expected background for massive binary black-holes (BBH) and galactic white-dwarf (WD) binaries \cite{Moore:2014lga}.}
\label{fig:Sensitivity}
\end{figure*}
 
 For GWs propagating in FLRW backgrounds, a source is present $\Pi_{ij}\neq0$ when there are additional tensor modes propagating. A paradigmatic example of this is bigravity, where there are two dynamical metrics. In that case, we have to track the evolution of both metric perturbations \cite{Narikawa:2014fua,Max:2017flc,Max:2017kdc}
 \be \label{eq:bigravity}
 \begin{pmatrix} h'' \\ t'' \end{pmatrix}+\lb k^2+m_g^2\begin{pmatrix} \sin^2\theta & -\sin\theta\cos\theta \\ -\sin\theta\cos\theta & \cos^2\theta \end{pmatrix}\rb\begin{pmatrix} h \\ t \end{pmatrix}=0\,,
 \ee
 where for shortness we have absorbed the Hubble friction in the definition of the perturbation and we do not show the spatial indices.
 Here $m_g$ is the effective mass (one of the tensor fields is massive) and $\theta$ is the mixing angle. Since there are interactions between $h_{ij}$ and $t_{ij}$, this means that the mass eigenstates are not the same as the propagation eigenstates. In analogy with the propagation of neutrinos, there can be GW oscillations. In Sec. \ref{sec:GWoscillations} we will see how GW oscillations can be tested. One should note that the possibility of having GW oscillations is not restricted to bigravity. Any gravity theory in which the additional degrees of freedom can arrange to form a tensor perturbation over FLRW background could display the same phenomenology. In particular, this is what happens with gauge fields in a SU(2) group~\cite{Caldwell:2016sut,BeltranJimenez:2018ymu}. 
 
 \subsection{Present and future GW detectors}
 \label{sec:Detectors}
 
Before presenting the different tests of gravity with multi-messenger GW astronomy, let us outline briefly the status of present and future GW detectors. We summarize the different sensitivities of each detector and the typical sources in Fig.~\ref{fig:Sensitivity}. The capabilities of multi-messenger GW astronomy depend mainly on two aspects:
\begin{itemize}
\item \emph{Number of detections:} this is most sensitive to the size of the volume of the Universe covered by the GW detector. However, there is a large uncertainty in the actual population of the sources, e.g. BNS. 
\item \emph{Sky localization:} this is most sensitive to the number of detectors that allow for a better triangulation of the source. A better localization of the GW events simplifies the search for a counterpart.
\end{itemize}
We draft a summary of present expectations for the range of detection and localization angle of different GW detectors in Fig.~\ref{fig:MMtimeline}. The reader should be aware that these expectations, specially the ones far in the future, might be subject to important modifications.

At present, we are in the second generation (2G) of ground-based detectors. There have been already two operation runs. In the first one, only the two aLIGO detectors were online with a detection range for BNS of the order of 80 Mpc. In the second one, aVirgo joined. Although its sensitivity was still lower, aVirgo helped to reduce the localization area an order of magnitude, from $100-1000\,\text{deg}^2$ to $10-100\,\text{deg}^2$. For illustration, we plot in Fig.~\ref{fig:Sensitivity} the strain of the first event GW150914 \cite{Vallisneri:2014vxa}.

However, neither aLIGO nor aVirgo has reached their designed sensitivity yet. Moreover, other two 2G detectors are on the way. KAGRA \cite{Somiya:2011np} in Japan is under construction and it is expected to start operating in 2020. On the other hand IndIGO \cite{LIGOIndia}, a replica of LIGO located in India has been approved. This means that in the coming years two main improvements are expected: a larger event rate and a more precise localization \cite{Aasi:2013wya}. The range of detection is expected to improve by a factor of 3 implying a factor 27 in the detection rate. The localization is expected to reduce to areas of $5-20\,\text{deg}^2$ with KAGRA and to a few $\text{deg}^2$ with IndIGO. Note that this is a key point in order to associate any counterpart with a GW event.

A third generation (3G) of ground-based detectors is being planned. The European 3G proposal is the Einstein telescope (ET) \cite{Sathyaprakash:2012jk}, an underground, three 10km-arms detector. Its current design aims at improving by a factor of 10 present sensitivity. The US 3G proposal, Cosmic Explorer (CE) \cite{Evans:2016mbw}, is more ambitious with two 40km arms further improving the sensitivity of ET. In any case, 3G detectors imply a substantial change in GW astronomy. While 2G detectors will only be able to reach up to $z\sim0.05$ for BNS and $z\sim0.5$ for BBHs, 3G detectors might reach $z\sim2$ for BNS and $z\sim15$ for BBHs. In terms of multi-messenger events, this corresponds to thousands or tens of thousands standard sirens.

The sky localization of events in 3G will vary depending on the available network of detectors \cite{Mills:2017urp}. In this sense, there are already plans to upgrade advanced LIGO detectors. This envisioned upgrade is known as LIGO Voyager \cite{LIGOplans}. Voyager could reach sensitivities between 2G and 3G. The localization will thus vary depending on the redshift of the source since the sensitivity of the network will not be homogeneous. A network of three Voyager detectors plus ET would localize $20\%$ of the events within $10\,\text{deg}^2$, while a setup with three ET detectors would localize $60\%$ of the events within $10\,\text{deg}^2$  \cite{Mills:2017urp}.
 
Moreover, space-based GW detectors have been also projected. The European space agency has approved LISA \cite{Audley:2017drz}. Being in space and with million kilometer arms, the frequency band and targets of LISA are very different from ground-based detectors (see Fig. \ref{fig:Sensitivity}). Expected sources include supermassive BHs, extreme mass ratio inspirals (EMRI) and some already identified white dwarf binaries (known as verification binaries). It is presumed that these sources could be observed with counterparts, enlarging the reach of multi-messenger GW astronomy. For reference, we have included in Fig.~\ref{fig:Sensitivity} the expected background of massive BBH ($M_{_\text{BH}}\sim10^{4-7}M_\odot$) and unresolved galactic white-dwarf binaries \cite{Moore:2014lga} (see more details about the different sources in Fig.~1 of \cite{Audley:2017drz}).
 
Finally, there are other proposals to detect GWs at even lower frequencies, in the band of 1-100 nHz. Sources in this regime could be binary SMBH in early inspiral or stochastic, cosmological backgrounds. These GWs could be observed using a network of millisecond pulsars, in which the pulsation is extremely well-known, for instance with PPTA \cite{Zhu:2014rta}. Other proposals are to use astrometry with GAIA, which is capable of tracking the motion of a billion stars \cite{Moore:2017ity}, or to use radio galaxy surveys \cite{Raccanelli:2016fmc}.  
  
\section{Standard sirens}
\label{sec:StandardSirens}

GWs coming from distant sources can feel the cosmic expansion in the same way as EM radiation does. In fact, we have seen in Sec. \ref{sec:GWcosmology} that the amplitude of the GWs is inversely proportional to the GW luminosity distance $d_{L}^{\text{gw}}$. In GR the GW luminosity distance is equal to EM luminosity distance, with the standard formula given by (\ref{eq:luminositydistance}).  However, this is not a universal relation in theories beyond GR as we will discuss in Sec. \ref{sec:GWdamping}. For the moment, we will restrict to Einstein's theory only.

\begin{figure}
\centering 
 \includegraphics[width=0.99\columnwidth]{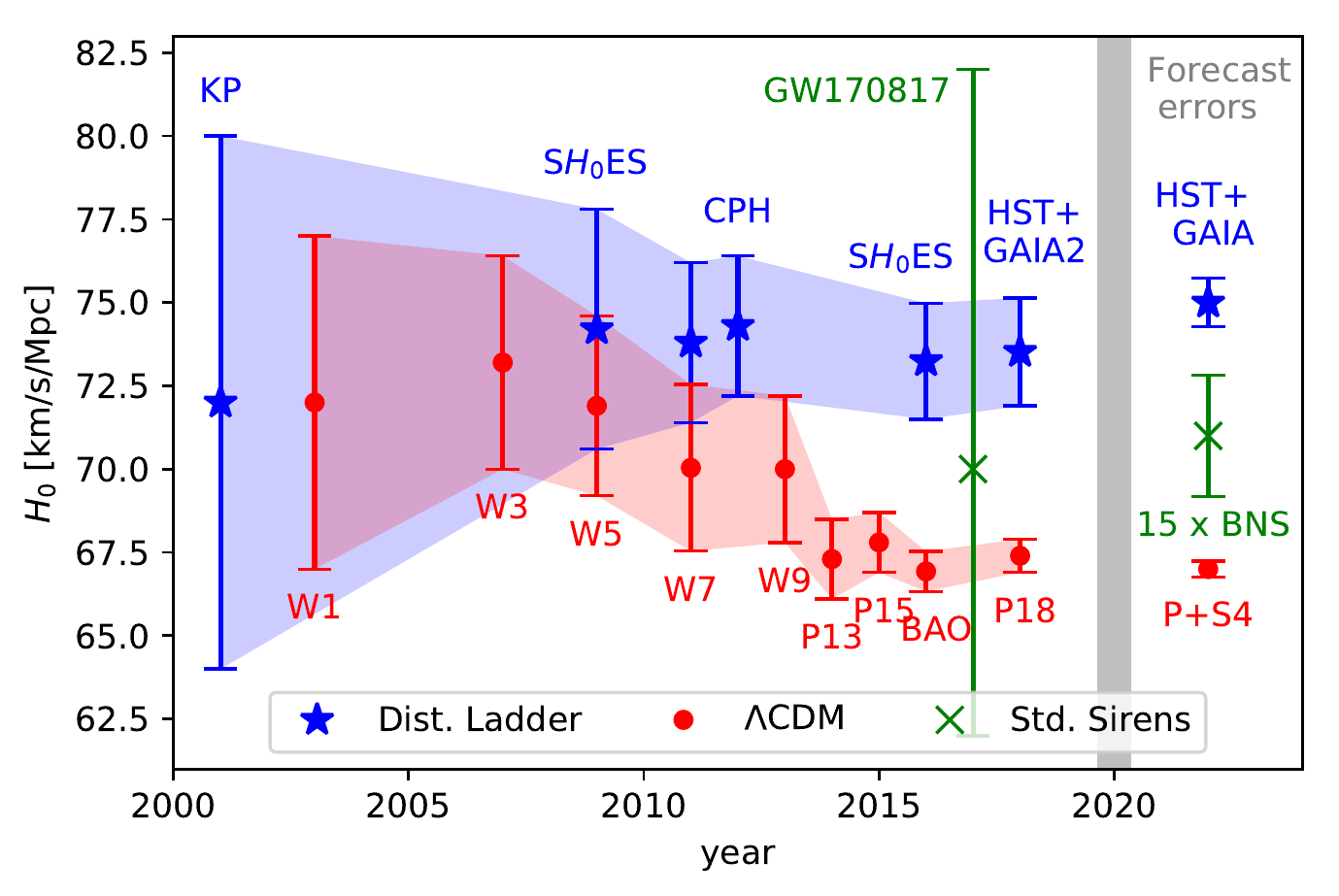}
 \caption{The Hubble tension (adapted from \cite{Freedman:2017yms,Beaton:2016nsw}, including the first standard sirens measurement following GW170817 \cite{Abbott:2017xzu}, Planck 2018 \cite{Aghanim:2018eyx} and Hubble Space Telescope (HST) with GAIA DR2 \cite{Riess:2018byc}). Blue stars correspond to measurements of $H_0$ in the local universe with calibration based on Cepheids. Red dots refer to derived values of $H_0$ from the CMB assuming $\Lambda$CDM. Green crosses are direct measurements of $H_0$ with standard sirens. Forecasts are: CMB Stage IV \cite{Abazajian:2016yjj}, standard sirens \cite{Nissanke:2013fka} and distance ladder with full GAIA and HST \cite{Casertano:2015dso,Riess:2016jrr}.
 }
 \label{fig:Hubbletension}
\end{figure}

In order to measure distances in cosmology one needs both a time scale and a proper ruler. The inverse dependence of the strain with $d_{L}^{\text{gw}}$ makes GWs natural cosmic rulers. Introducing the full cosmological dependence\footnote{In (\ref{eq:luminositydistance}) we had assumed a flat universe.}, the GW luminosity distance is given by
\be
\label{eq:luminositydistanceCosmo}
d_{L}^{\mathrm{gw}}=\frac{(1+z)}{\sqrt{\vert\Omega_K\vert}}\text{sinn}\lb c\int_0^z \frac{\sqrt{\vert\Omega_K\vert}}{H(z')}dz'\rb\,,
\ee
where $\text{sinn}(x)=\sin(x),\,x\,,\sinh(x)$ for a positive, zero and negative spatial curvature respectively. Assuming a $\Lambda$CDM cosmology, the Hubble parameter is a function of the matter content $\Omega_m$, the curvature $\Omega_K$ and the amount of DE $\Omega_\Lambda$ (radiation at present time is negligible)
\be
H(z)=H_0\sqrt{\Omega_m(1+z)^3+\Omega_K(1+z)^2+\Omega_\Lambda}\,.
\ee
On the contrary, GWs alone do not provide information about the source redshift. This is because gravity cannot distinguish a massive source at large distances with a light source at short distances. Nevertheless, when GWs events are complemented with other signals that allow a redshift identification, these events become \emph{standard sirens} \cite{Schutz:1986gp}. Standard sirens are complementary to already well-established  \emph{standard candles}, SN events in which the intrinsic luminosity can be calibrated allowing for a measurement of the EM luminosity distance. There are also \emph{standard rulers}, such as the one determined by baryon acoustic oscillations (BAO) which provides the angular diameter distance. For binary black-holes (BBH) it is not expected to observe any counterpart, unless there is matter around the BHs \cite{Perna:2016jqh}. Fortunately, binary neutron stars (BNS) and black-hole neutron star systems (BHNS) are expected to emit short gamma-ray bursts (sGRB) and other EM counterparts, becoming clear standard siren targets.

The first ingredient for a standard siren is the measurement of the GW luminosity distance. However, $d_{L}^{\text{gw}}$ is degenerate with the inclination of the binary. More precisely, showing the explicit angular dependence of the waveform (\ref{eq:WaveForm}) one finds that the two polarizations scale as
\be
h_+\propto\frac{(1+\cos\iota)^2}{2d_{L}^{\text{gw}}} \quad \text{and} \quad h_\times\propto \frac{\cos\iota}{d_{L}^{\text{gw}}}\,,
\ee
where $\iota$ is the inclination angle. This distance-inclination degeneracy is the main source of uncertainty of present measurements of $d_{L}^{\text{gw}}$ \cite{TheLIGOScientific:2016wfe}. One possibility to break this degeneracy is to have an identification of both polarizations. This requires at least a three detector network and a good sky localization. Another possibility to break this degeneracy is when the binary has a precessing spin. Then, there is a characteristic modulation of the amplitude that can disentangle the inclination angle. Orbital precession is more significant for large effective spin $\chi_\text{eff}$\footnote{The effective spin is the mass weighted projection of the two spins of the binaries into the orbital angular momentum.} and/or small mass ratios $q=m_2/m_1\leq1$ since there is also an effective spin-mass ratio degeneracy. Possibly good candidates for this would be BHNS binaries since BNS typically have a mass ratio close to 1.

The other ingredient for a standard siren is the identification of the redshift. This can be achieved by different means. The simplest consists in finding an EM counterpart of the GWs from the binary coalescence \cite{Schutz:1986gp}. Then, the redshift could be extracted from the EM counterpart or from the host galaxy depending on the case. BNS will produce a sGRB after the merger. This sGRB is characterized by a beaming angle $\theta_j$, which is typically expected to be $\theta_j\leq30^\circ$. This means that depending on the orientation of the source we will be able to detect both signals only in a small fraction of the events. Observing a bright afterglow or kilonovae \cite{Metzger:2016pju} might increase the changes of detecting a counterpart. BNS will be the primary source for LIGO \cite{Dalal:2006qt}, although BHNS could also play an important role \cite{Vitale:2018wlg}. SMBHs might be good standard sirens for LISA as well \cite{Holz:2005df}. Several multi-messenger observations will lead to a precise measurement of the cosmic expansion either for second generation detectors \cite{Nissanke:2009kt,Nissanke:2013fka} or for third generation \cite{Sathyaprakash:2009xt}. 

There are alternative proposals to identify the redshift without observing a counterpart. Based on statistical methods, one could associate every GW event with all the galaxies within the error in the localization and compute the cosmology \cite{Schutz:1986gp,DelPozzo:2011yh}. For a large number of events, the true cosmology will statistically prevail. Conveniently, this method applies to any type of source, including BBH which is the most common observation. Moreover, for very loud (golden) events there might be only few galaxies in the localization box \cite{Chen:2016tys}. On the con side, this method relies on a complete galaxy catalogue.

\begin{figure}[t!]
\centering 
 \includegraphics[width=0.99\columnwidth]{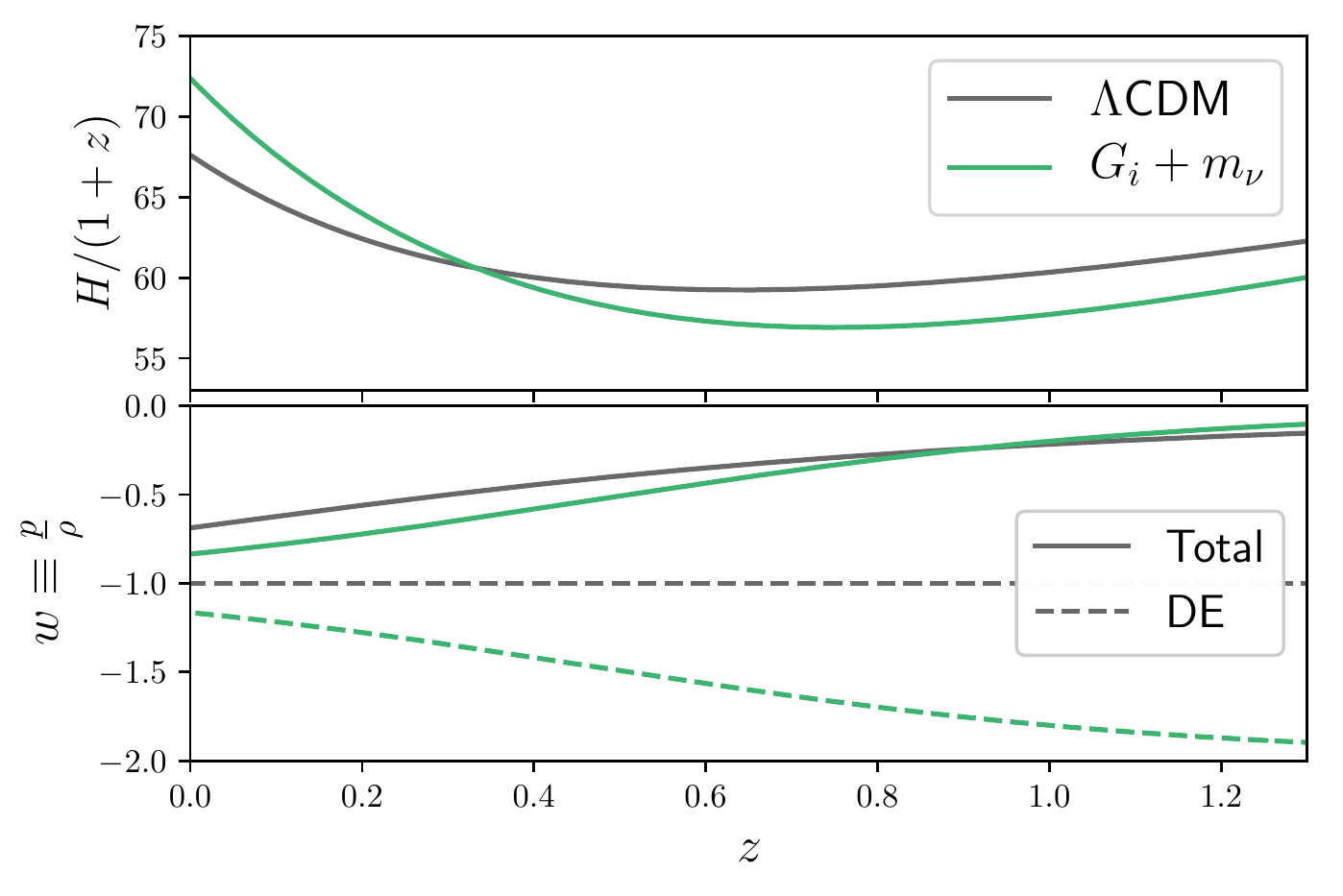}
 \caption{Hubble parameter $H$ and equation of state (EoS) $w$ as a function of redshift for the SM of Cosmology ($\Lambda$CDM) and for covariant Galileons with massive neutrinos. In the bottom panel, the total EoS $w_\text{tot}=p_\text{tot}/\rho_\text{tot}$ is compared with the EoS of DE $w_\text{DE}=p_\text{DE}/\rho_\text{DE}$.}
 \label{fig:H0eos}
\end{figure}

For events involving a NS there are other possibilities. If the EoS of the NS is known, one could compute the tidal effects in the GW phase, which breaks the degeneracy between the source masses and the redshift \cite{Messenger:2011gi}. A good sensitivity could be achieved with the Einstein Telescope \cite{DelPozzo:2015bna}. Since this method relies on the knowledge of the EoS, which most probably will be uncovered through GW observations also, an iterative approach could be performed. In addition, one could benefit from the narrow mass distribution of NS to statistically infer the redshift \cite{Taylor:2011fs}. Finally, numerical simulations suggests that in BNS a short burst of GWs with a characteristic frequency will be emitted after the merger. If this burst was observed, a redshift measurement could be obtained \cite{Messenger:2013fya}. The main challenge of this method is possibly the low SNR of the GW burst.

\begin{figure*}
\centering 
 \includegraphics[width=0.99\columnwidth]{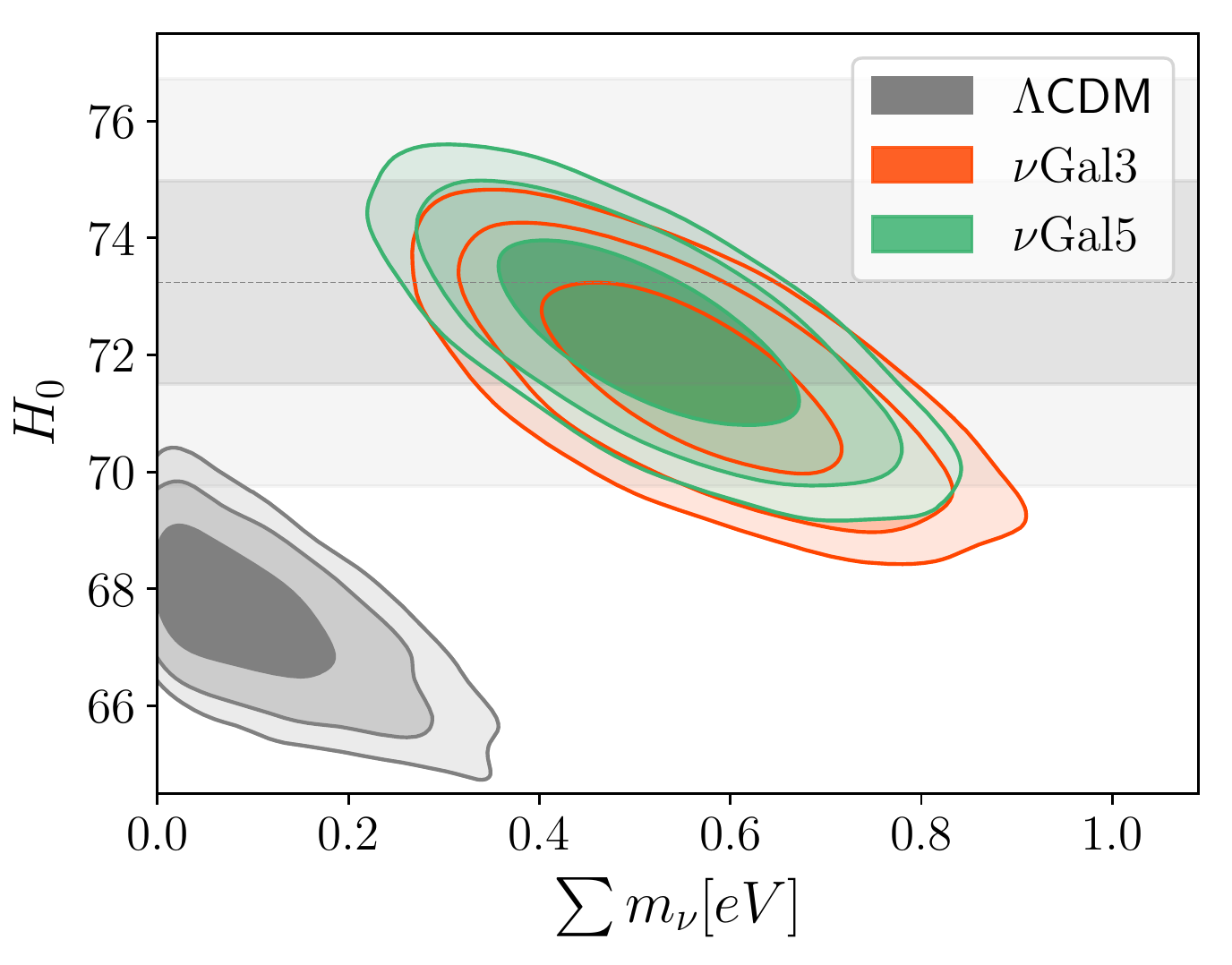}
  \includegraphics[width=0.99\columnwidth]{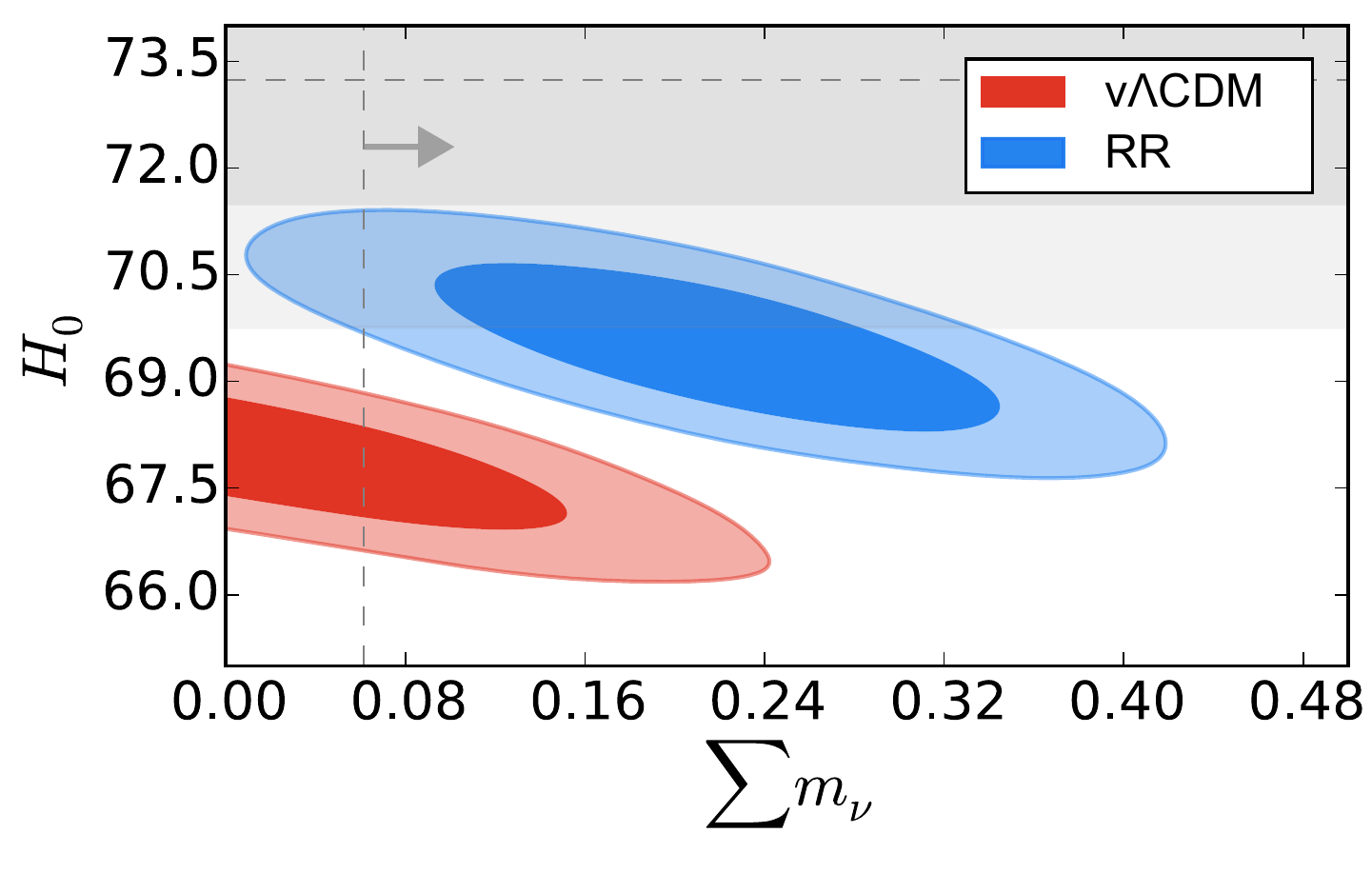}
 \caption{Fit of modified gravity models to Planck + BAO, marginalized over the Hubble constant $H_0$ and neutrino masses $\sum_\nu m_\nu$ for covariant Galileons (left) and for the non-local, RR-model (right). 
 A phantom-like equation of state $w<-1$ helps to solve the tension between Planck and the direct measurement (the non-zero neutrino mass partly compensates the effect of $w$, cf. Fig. \ref{fig:H0eos}).
 Figures reproduced with permission from the authors of \cite{Renk:2017rzu} and \cite{Belgacem:2017cqo} respectively. 
 \\
 \textcopyright\, SISSA Medialab Srl.. Reproduced by permission of IOP Publishing. All rights reserved.}
 \label{fig:H0inMG}
\end{figure*}

GW170817 has become the first standard siren detected. The redshift, $z=0.008^{+0.002}_{-0.003}$, was obtained identifying the host galaxy NGC4993 through the different EM counterparts \cite{GBM:2017lvd}. For such a close event, only the leading term in the cosmic expansion $H_0$ could be obtained following (\ref{eq:luminositydistanceExpansion}). The precise value obtained was~\cite{Abbott:2017xzu}, 
\be
H_0=70.0^{+12.0}_{-8.0}\text{km}\,\text{s}^{-1}\text{Mpc}^{-1}\,.
\ee
This result has the relevance of being the first independent measurement of $H_0$ using GWs. Still, since it is only one event, the relative error is large, of the order of $14\%$. From this error budget, $11\%$ arises from the uncertainty in the measurement of the distance due to present detector sensitivity and the previously mentioned degeneracy with the inclination angle. The rest of the error comes from the uncertainty in the estimation of the peculiar velocity of the host galaxy. Observations of the afterglow in different frequencies can help in reducing the inclination uncertainty \cite{Guidorzi:2017ogy,Hotokezaka:2018dfi}. One could also use the statistical method to obtain $H_0$ without information of the counterpart, although the error is significantly larger $H_0=76^{+48}_{-23}\text{km}\,\text{s}^{-1}\text{Mpc}^{-1}$ \cite{Fishbach:2018gjp}. Recent studies have shown that with order $\sim50$ BNS standard sirens events $H_0$ could be measured at the level of $\sim2\%$ \cite{Chen:2017rfc,Feeney:2018mkj}. Depending on the actual population of BNS this might be achieved with second generation detectors. LISA will detect mergers of SMBHs (with EM counterparts), providing measurements of cosmic expansion up to $z\sim8$ and potentially measuring $H_0$ with $0.5\%$ precision \cite{Tamanini:2016zlh}.

\subsection{The Hubble rate tension}
\label{sec:Hubbletension}
Standard siren observations of the cosmic expansion can also explore the tension on the Hubble parameter: where a distance ladder measurement gives a value $H_0 = (73.52\pm 1.62) {\rm km}\, s^{-1}{\rm Mpc}^{-1}$ \cite{Riess:2018byc} higher than the model-dependent inference from the CMB $H_0 = (67.4 \pm 0.5) {\rm km}\, s^{-1}{\rm Mpc}^{-1}$ \cite{Aghanim:2018eyx} (see in Fig. \ref{fig:Hubbletension}). The tension now reaches the level of $3.6\sigma$. 
Reanalysis of the local distance ladder with more sophisticated statistical techniques tend to agree on the high value, although with somewhat larger error bars \cite{Cardona:2016ems,Feeney:2017sgx}.
Other low redshift determinations confirm this trend, for instance time delays from multiply-imaged quasar systems \cite{Bonvin:2016crt} give $H_0 = (71.9^{+2.4}_{-3.0}) {\rm km}\, s^{-1}{\rm Mpc}^{-1}$.  
Measurements of $H_0$ can also be obtained combining BAO and primordial deuterium abundances \cite{Addison:2017fdm} (see more details in the review \cite{Suyu:2018vqs} and a compilation of the values of $H_0$ in~\cite{Bernal:2018cxc}).

If the tension is not due to systematic errors in either of the surveys, it would indicate a mismatch between the low and high redshift distance ladders \cite{Cuesta:2014asa}, which might be the first hint of the need to revise the standard cosmological model.
Several partial solutions to the $H_0$ tension have been proposed, although no satisfactory solution exists. Extensions to $\Lambda$CDM have been studied, but no simple model seems to work: for instance, increasing the effective number of relativistic species by $\Delta N_{\rm eff} \approx 0.4$ eases the tension but enters in conflict with small scale Planck polarization \cite{Bernal:2016gxb}, which has been confirmed in the latest Planck results. 
The role of dark energy (through $w(z)$) has also been investigated in connection with the $H_0$ tension: no equation of state evolution $w(z)$ can reconcile all datasets, as long as GR holds (although the tension could be eased if BAO or SNe data are not included) \cite{Poulin:2018zxs}.
Interacting DE eases the tension, particularly for a phantom-like equation of state with $w\sim -1.2$ \cite{DiValentino:2017iww}.

Some dark energy models beyond GR and with massive neutrinos have been proposed to ease the tension. 
Galileon gravity leads to a phantom-like equation of state (EoS) $w<-1$ \cite{DeFelice:2010pv,Barreira:2013jma}: 
adding massive neutrinos with total mass $\sum m_\nu \approx 0.6 eV$ yielded a good fit to both Planck and the direct $H_0$ measurement \cite{Barreira:2014jha}. One should note that although the EoS of Galileons $w_\text{Gal}$ deviates significantly from $w_\Lambda=-1$, massive neutrinos compensate part of the effect so that the total EoS $w_\text{tot}=p_\text{tot}/\rho_\text{tot}$ is more similar to $\Lambda$CDM (see bottom panel of Fig. \ref{fig:H0eos}). Still, this difference is enough to shift the present value of the Hubble parameter $H_0\equiv H(z=0)$ to higher values (see upper panel of Fig. \ref{fig:H0eos}).   
A latter analysis, shown in Fig. \ref{fig:H0inMG}, reproduced the result, but found a slight tension with the most recent BAO data \cite{Renk:2017rzu}. Most importantly, the cosmologically viable Galileons were ruled out by GW speed \cite{Ezquiaga:2017ekz} and weak lensing \cite{Peirone:2017vcq}.
Note however that those data employed BAO reconstruction and Galileons are known to affect the non-linear BAO evolution \cite{Bellini:2015oua}, making it more conservative to use the unreconstructed data, for which no tension exists. 
Non-local gravity has similar features (cf. Fig. \ref{fig:H0inMG}) but its less negative equation of state (compensated with $\sum m_\nu \approx 0.3$) leads to a reduced tension rather than close agreement  \cite{Belgacem:2017cqo}.

\section{Gravitational wave speed}
\label{sec:GWspeed}

The speed of GWs is a fundamental property of any gravity theory. GR predicts that GWs propagate at the speed of light. However, alternative theories generically change this prediction. In contrast to (\ref{eq:dispersionGR}), GWs in modified gravity do not have to travel on null geodesics of the background metric. One can parametrize the generalized propagation by
\be
\mathcal{G}^{\mu\nu}k_\mu k_\nu+m_g^2+\sum_{i=3}^{n}A^{\alpha_1\cdots\alpha_n}k_{\alpha_1}\cdots k_{\alpha_n}=0\,.
\ee
Here, $\mathcal{G}^{\mu\nu}$ is the effective metric over which GWs propagate, $m_g$ is the effective mass of the graviton and the tensors $A^{\alpha_1\cdots\alpha_n}$ encode higher-order, wave-vector corrections. When time and space can be decomposed, the above expression leads to a generalized dispersion relation 
\begin{equation}
\label{eq:dispersion}
\omega^2 = c_g^2k^2 + m_g^2+\sum_{n=3}\mathbb{A}_{n}k^{n}\,,
\end{equation}
where $k$ is the spatial modulus of the wave-vector and $\mathbb{A}_{n}$ are the coefficients of the higher order corrections. Accordingly, we can see that the effective metric determines the propagation speed $c_g$ \cite{Bettoni:2016mij} while the higher order wave-vector corrections control Lorentz-violating modifications of the dispersion relation \cite{Mirshekari:2011yq}. The mass term $m_g$ also modifies the dispersion relation \cite{Will:1997bb}. In the following, we discuss the origin of and the constraints on these three different contributions. We will focus on constraints from late time GW sources. A modified dispersion relation for primordial GWs could be tested with the B-mode polarization of the CMB, as it has been studied for the case of the speed $c_g$ \cite{Amendola:2014wma,Raveri:2014eea,Pettorino:2014bka}, and the mass $m_g$ \cite{Dubovsky:2009xk,Brax:2017pzt,Fasiello:2015csa}.

\subsection{Anomalous GW speed}
\label{sec:AnomalousSpeed}

\begin{figure}[t!]
\centering
\includegraphics[width = 0.95\columnwidth,valign=t]{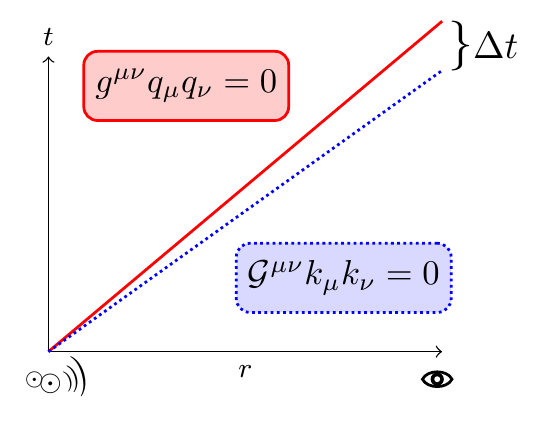}
\caption{Anomalous GW speed. Gravitational waves propagate on an effective metric $\mathcal{G}^{\mu\nu}$ (blue) with a different causal structure than the physical metric $g^{\mu\nu}$ (red) \cite{Bettoni:2016mij}. The speed is derived as $c_g(\vec k) = \omega(\vec k)/|\vec k|$ where $k^{\mu}=(\omega,\vec k)$ is the solution to $\mathcal{G}^{\mu\nu}k_\mu k_\nu=0$. Note that the speed can  depend on the propagation direction. It may also depend on the frequency (e.g. massive graviton or Lorentz violation), cf. (\ref{eq:dispersion}).}
\label{fig:GWcausality}
\end{figure}

In order to obtain the frequency independent propagation speed $c_g$, one has to focus on the leading derivative terms for the second order action for the tensor perturbations $h$. At small scales and for arbitrary backgrounds, the action is determined by the effective metric $\mathcal{G}^{\mu\nu}$ over which GWs propagate \cite{Bettoni:2016mij}
\be
\Lag\propto h_{\mu\nu}\mathcal{G}^{\alpha\beta}\partial_\alpha\partial_\beta h^{\mu\nu}=h_{\mu\nu}\lp \mathcal{C}\Box+\mathcal{D}^{\alpha\beta}\partial_\alpha\partial_\beta\rp h^{\mu\nu}\,.
\ee
The effective metric can be further decomposed in a piece proportional to the original metric $\mathcal{C}$ and another not proportional $\mathcal{D}$. Then, whenever the (non-conformal) second term is present, the \emph{GW-cone} will be different from the \emph{light-cone} and both signals will travel at different speeds (see Fig. \ref{fig:GWcausality}).\footnote{Note that similar arguments could be applied to the other gravitational modes, for instance for a scalar field \cite{Babichev:2007dw}.}

In scalar-tensor gravity, two conditions have to be fulfilled to induce an anomalous propagation speed: $i)$ there is a non-trivial scalar field configuration (if we want to explain DE, we typically demand $\dot{\phi}\sim H_0$) and $ii)$ there is a derivative coupling to the curvature. This highlights the presence of a modified gravity coupling that will lead $\mathcal{D}^{\alpha\beta}\sim\partial^\alpha\phi\,\partial^\beta\phi$. Whenever these two conditions are satisfied, $c_g\neq c$ and there would be a delay between the GW and the EM counterpart. For instance, differences of $1\%$, $c_g/c\sim0.01$, for sources at $100$Mpc induce delays of $\Delta t \sim10^7$years, clearly beyond human timescales.

Similar arguments can be applied to other gravity theories with additional degrees of freedom. Massive gravity and bigravity have a canonical kinetic term for the gravitons (due to the Einstein-Hilbert term) and thus GWs propagate at the speed of light. In vector-tensor theories there could be couplings to the curvature leading to an anomalous propagation speed, for instance $R_{\mu\nu}v^\mu v^\nu$ in vector DE \cite{Jimenez:2008au}. Interestingly, in more complex vector theories, it is possible to have derivative couplings to the curvature through the field strength $F^{\mu\nu}$ which do not induce an anomalous speed over cosmological backgrounds \cite{BeltranJimenez:2018ymu}. This is because in these theories it is possible to have cosmic acceleration while the background of $F^{\mu\nu}$ vanishes, thus violating condition $i)$. One should notice that, when violating some of the initial assumptions, the propagation speed of GWs might not be subject to the background value of any additional field and just to the parameters of the theories. This is the case for instance of Ho\v{r}ava gravity \cite{Blas:2014aca}.

Alternatively, a much more common strategy followed in the literature is to compute the speed of GWs directly in a given background, usually FLRW. For Horndeski theory this was done in \cite{Kobayashi:2011nu,Bellini:2014fua}. The implications of an anomalous GW speed have been discussed for instance for purely kinetic coupled gravity \cite{Kimura:2011qn}, covariant Galileons \cite{Brax:2015dma} and models with self-acceleration \cite{Lombriser:2015sxa,Lombriser:2016yzn}. The implications for cosmology were discussed in \cite{Saltas:2014dha,Sawicki:2016klv}. In vector-tensor theories, cosmological tensor perturbations have been computed for instance in~\cite{DeFelice:2016yws,DeFelice:2016uil}.

Prior to the direct detection of GWs, there were indirect constraints on the speed of GWs. High energy cosmic rays from galactic origin set a stringent lower bound $-2\cdot10^{-15}\leq c_g/c-1$ \cite{Moore:2001bv}, due to the absence of gravitational Cherenkov radiation \cite{Caves:1980jn}. The reason is that if gravitons propagate slower than the speed of light, cosmic rays could decay into them and their signal would be lost. This lower bound affects Horndeski theory \cite{Kimura:2011qn}. However, note that we are talking about very energetic gravitons, different from the low energy GW emission of an astrophysical compact binary. Moreover, the GW speed was indirectly constrained at the level of $\vert c_g/c-1\vert \leq 0.01$ with the orbits of binary pulsar in the absence of screening of the cosmological solution \cite{Jimenez:2015bwa}.  

With the detections of GWs from BBHs, the first direct constraints on the speed of GWs were placed \cite{Blas:2016qmn,Cornish:2017jml}. The constraints were still not very strong, $-0.45\leq c_g/c-1\leq0.42$, due to the uncertainties in the localization of the source and the low number of detections (3 at the time of the analysis). Detecting a GW with an EM counterpart changes the situation completely, leading to very precise measurements \cite{Will:1997bb,Nishizawa:2014zna,Nishizawa:2016kba} 

Such a multi-messenger GW event was detected on August 17, 2017 with the BNS GW170817 \cite{TheLIGOScientific:2017qsa}. The GW signal was followed by a short gamma ray burst (sGRB) only $\Delta t=1.74\pm0.05$s after \cite{Monitor:2017mdv}. The source was localized at a distance of $d_{_\text{L}}=40^{+8}_{-14}$Mpc. In order to set the constraints, the LIGO-Virgo collaboration conservatively considered the source at the lowest distance of 26Mpc. For the upper bound, it was assumed that both the GW and the sGRB were emitted at the same time and that all the delay is caused by the faster propagation of the GW. For the lower bound, they assumed that the sGRB was generated 10s after the GW, order of magnitude expected in standard astrophysical models, and that the delay was reduced to 1.74s due to the slower propagation of the GW. In total, this led to the impressive constraint
\be
-3\cdot 10^{-15}\leq c_g/c-1\leq 7\cdot 10^{-16}\,.
\ee
This result has profound implications for many gravity theories and dark energy models.

In scalar-tensor gravity at least one of the conditions for an anomalous GW speed has to be broken. If we want the scalar field to keep playing a role in the cosmic expansion history, it cannot have a trivial scalar field configuration. Therefore, the only possibility to satisfy GW170817 is to break the second condition an eliminate derivative couplings to the curvature. For Horndeski theory (\ref{eq:LH2}-\ref{eq:LH5}) this implies \cite{Ezquiaga:2017ekz,Creminelli:2017sry,Baker:2017hug,Sakstein:2017xjx}
\begin{equation}
G_{4,X} \approx 0 \,,\quad G_5 \approx \text{constant}\,.
\end{equation}
Translating this result, only the simplest models such as quintessence, Brans-Dicke or Kinetic Gravity Braiding survive. On the contrary, models like Covariant Galileons, Fab Four, Gauss-Bonnet or some sectors of beyond Horndeski are ruled out. The fact that the parameter space has been drastically reduced has implications for cosmological constraints \cite{Kreisch:2017uet,Peirone:2017ywi,Arai:2017hxj} and for large scale structure \cite{Amendola:2017orw}. 

For vector-tensor theories the situation is very similar. In order to describe DE and to pass the GW test some couplings of the theory have to be eliminated \cite{Ezquiaga:2017ekz,Baker:2017hug}, in particular $G_{4,Y} \approx 0$ and $G_{5,Y} \approx 0$ (see full action in Eq. (299) of \cite{Heisenberg:2018vsk})
The same happens for Ho\v{r}ava gravity where one has to impose $\xi \approx 1$ or $\beta_{kh}\approx 0$ \cite{Gumrukcuoglu:2017ijh},
which correspond to the conditions for the low-energy version of the theory or its Einstein-aether analogue respectively. The implications of GW170817 for other gravity theories have been extensively explored, for instance for
 doubly-coupled bigravity \cite{Akrami:2018yjz}, $f(T)$ gravity \cite{Cai:2018rzd} or Born-Infeld models \cite{Jana:2017ost}. 

\subsection{Mass term}

A graviton mass, either effective or fundamental, modifies the propagation speed of GWs. However, contrary to the anomalous speed term $c_g$, it does it in a frequency dependent way. This means that it can be constrained with GW observation alone, analyzing how the phase of the wave changes in time. The present bound from the LIGO-Virgo collaboration is \cite{Abbott:2017vtc}
\be
m_g\leq7.7\cdot10^{-23}\,\text{eV}/c^2\,.
\ee 
Note that this bound is still far away from the cosmologically "motivated" $m_g\sim H_0\simeq10^{-33}\text{eV}/c^2$.

Since a graviton mass would also change gravity in other regimes, we can compare the GW bound with other tests. In particular, a massive graviton introduce a Yukawa potential that can be constrained with Solar System observations. This issue has been recently revisited \cite{Will:2018gku}, showing that the best bound comes from the perihelion advance of Mars, leading to $m_g<(4-8)\cdot10^{-24}\text{eV}/c^2$, which is an order of magnitude better than present GW constraints. 

For LISA, the GW bound could improve significantly, due to the lower frequencies and higher distances, possibly reaching $m_g<10^{-26}\text{eV}/c^2$ \cite{Berti:2004bd}. In addition, there are proposals to bound $m_g$ measuring the phase lag of continuos sources of GWs and EM radiation with LISA binaries \cite{Larson:1999kg,Cutler:2002ef,Finn:2013toa}.\footnote{In fact, one can use the phase lag test to constraint the propagation speed of GWs in general \cite{Bettoni:2016mij}.} For more details in other types of constraints, we recommend the recent review \cite{deRham:2016nuf}. 

\subsection{Modified dispersion relation}

Similarly to a graviton mass, Lorentz violating terms modify the dispersion relation in a frequency dependent way. Different wavelengths thus travel at different speeds, modifying the time evolution of GW phase. The effects of the new terms $\mathbb{A}_{i}$ in the dispersion relation (\ref{eq:dispersion}) can be systematically parametrized in modifications of the waveform \cite{Mirshekari:2011yq}. A typical example of a Lorentz-violating theory would be high-energy Ho\v{r}ava gravity \cite{Horava:2009uw} in which
\be
\omega^2=c^2k^2+\frac{\kappa^4_{h}\mu_{h}^2}{16}k^4+\cdots\,,
\ee
where $\kappa_{h}$ and $\mu_{h}$ are parameters of the theory \cite{Vacaru:2010rd}.

From the first two events, GW150914 \cite{Abbott:2016blz} and GW151226 \cite{Abbott:2016nmj}, one can already constraints several theories as detailed in Ref. \cite{Yunes:2016jcc}. For Ho\v{r}ava gravity, one can constrain the combination of parameters $\kappa^4_{h}\mu_{h}^2$, which were not bounded previously. GW170104 \cite{Abbott:2017vtc} and GW170817 \cite{Monitor:2017mdv} have also been used by LVC to constrain the different $\mathbb{A}_{n}$.

\subsection{Equivalence principle}

The fact that GWs and EM radiation from GW170817 arrived almost simultaneously at Earth after approximately 100 million light years of travel tells us that both signals follow very similar geodesics. This statement can be made precise in terms of the Shapiro delay \cite{Shapiro:1964uw}. The Shapiro delay measures the difference on arrival time of a massless particle in flat and curved space-time. This can be computed parametrizing the integral of the gravitational potential $U(\mathbf{r})$ over the line of sight \cite{Krauss:1987me}
\be
\Delta t_S=-\frac{(1+\gamma)}{c^2}\int_{\mathbf{r}_{e}}^{\mathbf{r}_{o}} U(\mathbf{r}(l))dl\,,
\ee
where $\mathbf{r}_{e}$ and $\mathbf{r}_{o}$ are the positions at emission and observation. The prediction of GR is that $\gamma=1$ for any massless particle. This has been tested to very good precision for photons, $\gamma_\text{em}-1\leq(2.1\pm2.3)\cdot10^{-5}$, using the Cassini space-craft \cite{Bertotti:2003rm}. This is one of the most stringent Solar System test of gravity and implies that in these scales the gravitational potential should be very similar to GR as discussed in detail in the review \cite{Will:2014kxa}. 

Now, the multi-messenger observation of GW170817 allow us to test if GWs and EM radiation feel the same gravitational potential. In other words, this is testing the equivalence principle. In order to get a bound on the relative difference of $\gamma_\text{gw}$ and $\gamma_\text{em}$ one needs to know the gravitational potential between the BNS and the detectors. A conservative bound can be placed introducing only the effect of the Milky Way to arrive at \cite{Monitor:2017mdv}
\be
-2.7\cdot 10^{-7}\leq \gamma_\text{gw}-\gamma_\text{em}\leq 1.2\cdot 10^{-6}\,.
\ee
This constraint has implications for instance for theories in which the dark matter arises from a non-minimal matter coupling to gravity, the so-called dark matter emulators \cite{Boran:2017rdn}. If both types of waves propagate in the same effective metric, no relative difference is present, as it has been argued for the case in MOG gravity \cite{Green:2017qcv}.

\section{Gravitational wave damping}
\label{sec:GWdamping}

Apart from the speed of GWs, the other main observable from the modified propagation is the luminosity distance of GWs $d_L^{\text{gw}}$. For theories in which $c_g=c$, the GW luminosity distance (\ref{eq:MGluminosity}) is related to the EM luminosity distance $d_L^{\text{em}}$ by
\be
\frac{d_L^{\text{gw}}(z)}{d_L^{\text{em}}(z)}=\text{exp}\lb\frac{1}{2}\int_0^z  \frac{\nu}{1+z'}dz'\rb\,,
\ee
where $\nu$ is the additional friction term from modifying gravity, cf. (\ref{eq:ModProp}). Therefore, one can probe the damping of GWs using standard sirens, since for those multi-messenger observations both $d_L^{\text{gw}}(z)$ and $d_L^{\text{em}}(z)$ are measured \cite{Deffayet:2007kf}. Moreover, even without an EM counterpart, any additional friction for the GWs could be probed using GW source counts \cite{Calabrese:2016bnu}.


\begin{figure}
\centering 
 \includegraphics[width=0.99\columnwidth]{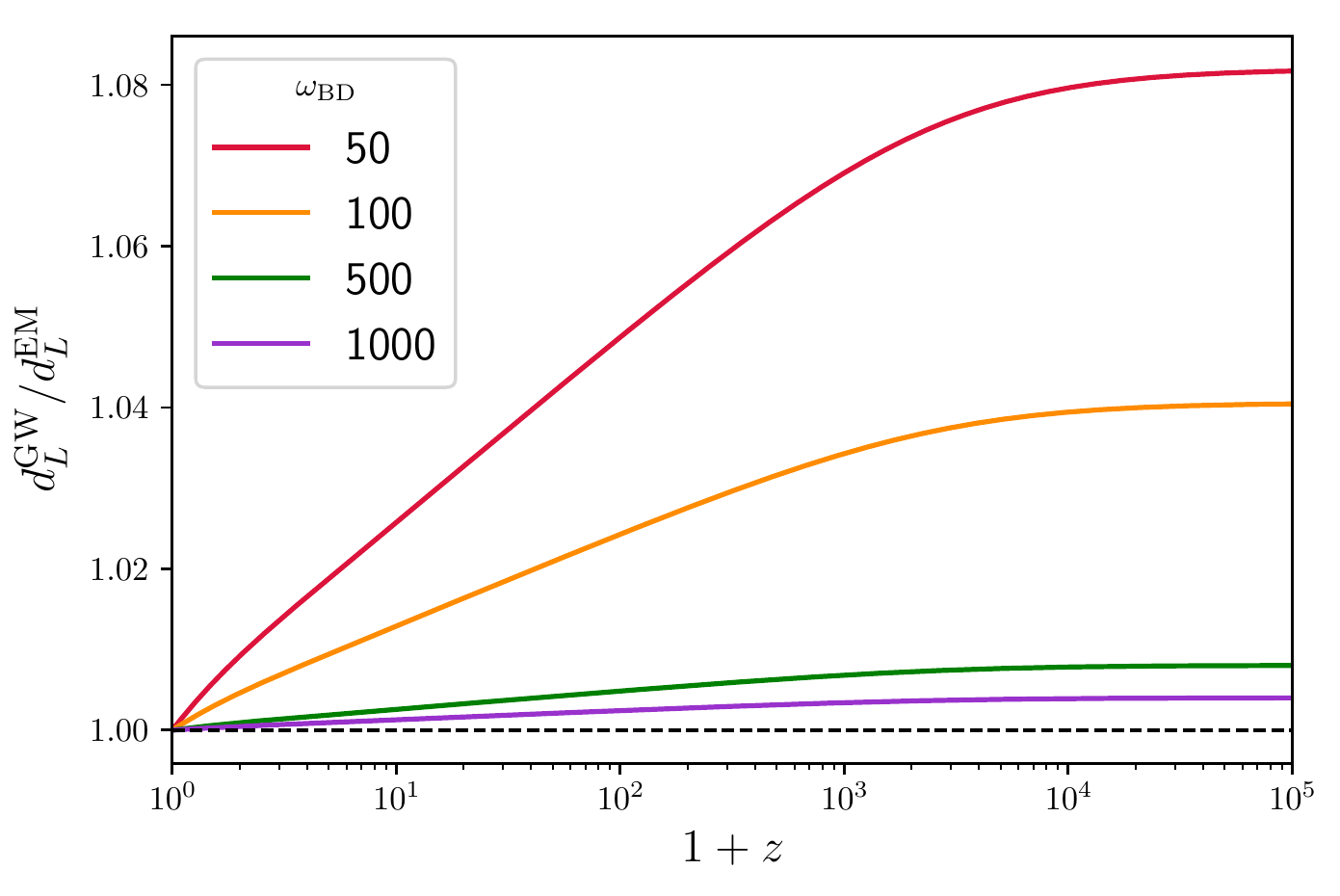}
 \caption{Ratio between the GW and the EM luminosity distances in Brans-Dicke for different values of $\omega_{BD}$, cf. Eq. (\ref{eq:lag_brans_dicke}).}
 \label{fig:GWluminosity}
\end{figure}

A paradigmatic example of a modification of gravity in which the GW luminosity distance differs from the EM one is adding extra dimensions \cite{Deffayet:2007kf}. In extra dimension theories, for instance DGP, there can be a large distance leakage of the gravitons into the additional dimensions. This means that, as a net effect, an observer will receive less gravitons or, in other words, the gravitational signal will be dimmer. By dimensional analysis, the GW luminosity distance scales in these theories as
\be \label{eq:dLextradim}
\frac{d_L^{\text{gw}}(z)}{d_L^{\text{em}}(z)}\propto\lp d_L^{\text{em}}(z)\rp^{(D-4)/2}\,,
\ee
where $D$ refers to the number of space-time dimensions in which the graviton can propagate.\footnote{For an analysis of the GW propagation over compact extra dimensions see Ref. \cite{Andriot:2017oaz}.} For $D=4$, one recovers the GR result $d_L^{\text{gw}}=d_L^{\text{em}}$. In cases in which the graviton can only travel in the extra dimensions above a certain screening scale $R_c$, the previous relation generalizes to \cite{Abbott:2018lct}
\be \label{eq:dLextradimscreen}
\frac{d_L^{\text{gw}}(z)}{d_L^{\text{em}}(z)}=\lb1+\lp\frac{d_L^{\text{em}}}{R_c}\rp^{n}\rb^{(D-4)/(2n)}\,,
\ee
where $n$ measures the transition steepness and the GR limit is recovered when $D=4$.

In scalar-tensor gravity it is also known how the GW luminosity distance will evolve. The additional friction is equal to the effective Planck mass run rate $\alpha_{_\text{M}}$
\be
\nu=\alpha_{_\text{M}}=\frac{d\ln M_*^2}{d\ln a}\,,
\ee
where $M_*$ is the effective Planck mass, i.e. the normalization of the kinetic term of the tensor perturbations. Then, recalling the redshift definition $1+z=a_0/a$, one arrives at
\be
\frac{d_L^{\text{gw}}(z)}{d_L^{\text{em}}(z)}=\frac{M_*(0)}{M_*(z)}\,,
\ee
where $M_*(0)$ and $M_*(z)$ are the effective Planck masses at the time of observation and emission respectively. Assuming that there is no screening and taking $\alpha_{_\text{M}}$ constant, one could rewrite this expression as \cite{Nishizawa:2017nef}
\be
\frac{d_L^{\text{gw}}(z)}{d_L^{\text{em}}(z)}=(1+z)^{\alpha_{_\text{M}}/2}\,.
\ee
For this case, the implications of measuring $\alpha_{_\text{M}}$ for Horndeski cosmology have been discussed in \cite{Saltas:2014dha,Lombriser:2015sxa}. The prospects of constraining the time variation of the Planck mass has been investigated for aLIGO in \cite{Nishizawa:2017nef} and for LISA in \cite{Amendola:2017ovw}. For illustration, we plot in Fig.~\ref{fig:GWluminosity} how the ratio $d_L^{\text{gw}}(z)/d_L^{\text{em}}(z)$ would vary in Brans-Dicke depending on the coupling $\omega_{BD}$, cf. Eq. (\ref{eq:lag_brans_dicke}).

Another theory in which the GW luminosity distance has been investigated is the non-local, RR-model. For this model, one finds \cite{Belgacem:2017ihm}
\be
\frac{d_L^{\text{gw}}(z)}{d_L^{\text{em}}(z)}=\sqrt{\frac{G_{\text{eff}}(z)}{G_{\text{eff}}(0)}}\,,
\ee
where the effective Newton constant is related to the parameters of the theory
\be
G_\text{eff}(z)=\frac{G}{1-3\gamma\bar{V}(z)}\,,
\ee
with $\bar{V}(z)$ being the background evolution of the auxiliary field and $\gamma=m^{2}/(9H_0^2)$ linked to the mass of the conformal mode $m$ (see details in the review \cite{Belgacem:2017cqo}). Thus, the growth of structure is directly related to the GW propagation. This behavior is also reproduced in some Horndeski models \cite{Linder:2018jil}. Differently to the scalar-tensor case, there is no screening for these non-local models. One should note also that the strength of the modification of the GW luminosity relation is sensitive to the initial conditions of the auxiliary field $\bar{V}(z)$, which depends on the unknown early universe physics.

With the detection of the multi-messenger event GW170817 it was possible to test the gravitational Hubble diagram for the first time. The observation was consistent with GR although being just one event the constraining power is still moderate. For theories with extra dimensions following (\ref{eq:dLextradim}), it was found that the number of space-time dimensions in which the gravitons propagate is limited to \cite{Pardo:2018ipy}
\be
D=4.02^{+0.07}_{-0.10}\quad \text{or}\quad D=3.98^{+0.07}_{-0.09}
\ee
for SN or CMB prior in $H_0$ (see Fig. \ref{fig:Hubbletension} and Sec. \ref{sec:Hubbletension}).\footnote{This model was reanalyzed in \cite{Abbott:2018lct} without assuming any prior in $H_0$ but the GW170817 measurement and including the screening~(\ref{eq:dLextradimscreen}), which differs from the one in \cite{Pardo:2018ipy}.} Similar analysis follows for brane-world models \cite{Visinelli:2017bny}, constraining in that case the radius of curvature of the extra dimensions. 
Moreover, the additional friction $\nu$ can only be loosely constrained \cite{Arai:2017hxj}
\be
-75.3\leq \nu\leq78.4\,.
\ee
In order to connect this result with the previously discussed theories recall that for scalar-tensor gravity $\nu=\alpha_{_\text{M}}$ and for the non-local, RR-model $\nu=-2\delta$ \cite{Belgacem:2017ihm}.

An important remark when evaluating the GW luminosity distance in modified gravity is that it will not only be altered with respect to GR due to the modified propagation of GWs but also because the cosmological expansion history is different. In other words, in alternative theories of gravity both the EM luminosity distance $d_L^{\text{em}}$ and its relation with the GW luminosity distance can be modified, due to a different $H(z)$ and to an additional friction $\nu$ respectively. In fact, the contribution of the the modified propagation can dominate over the modified cosmic expansion history. Introducing a phenomenological parametrization of the GW luminosity distance \cite{Belgacem:2018lbp}
\be
\frac{d_L^{\text{gw}}(z)}{d_L^{\text{em}}(z)}=\Xi_0+\frac{1-\Xi_0}{(1+z)^n}\,
\ee
together with the usual $(w_0,w_a)$ parametrization of $H(z)$, it was shown that the largest contributions are $\Xi_0$ and $w_0$. The prospects of measuring $\Xi_0$ with the Einstein telescope were also considered in \cite{Belgacem:2018lbp}. 

 \section{Additional polarizations}
 \label{sec:GWPolarization}
 
Apart from the modified GW propagation, the other main GW effect of theories beyond GR would be the emission of additional polarizations. We have seen that observing the orbits of pulsars already severely constrains the gravitational energy loss to that of GR. Now, GW astronomy enables to directly probe these extra modes. For this test, the basic role of multi-messenger events is improving the localization and breaking degeneracies with the orientation.
\footnote{In some sense, one could argue that a simultaneous detection of GR and non-GR polarizations is a multi-messenger observation itself.}
   
With direct GW observations, the emission of additional polarizations can be constrained from the modifications of the waveforms. For instance, with the first two events it was possible to limit the presence of scalar hair \cite{Yunes:2016jcc}. However, there are still degeneracies between the modified GW phase and the spin and mass parameters that weaken the constraints. This is the case of Einstein-dilaton-Gauss-Bonnet \cite{Sotiriou:2013qea} and dynamical Chern-Simons gravity \cite{Jackiw:2003pm}, archetypical examples of theories studied in numerical relativity \cite{Yagi:2016jml,Benkel:2016kcq}.
 
Moreover, there are also searches for direct signals of non-tensorial polarizations, analyzing the GW geometry through the projection of the different polarizations $A_{ij}$ (\ref{eq:polarizations}) onto the detector's network. Since the two LIGO interferometers Hanford and Livingston are basically coaligned, they maximize the SNR of the detection but are insensitive to polarizations. This situation changes with the incorporation of Virgo. From the observation of GW170814, a three detector BBH signal, pure tensor polarization were favored over pure vector or pure scalar modes \cite{Abbott:2017oio,Isi:2017fbj}. However, this was just a simplified analysis and the LIGO-Virgo collaboration is performing a more intensive study including mixed-polarization, which would be a more realistic setup. In the future, these constraints will improve with the switch on of the Japanese detector KAGRA and aLIGO India (see Fig.~\ref{fig:MMtimeline}). Nevertheless, one should note that quadrupolar detectors like aLIGO and aVirgo cannot distinguish between the breathing and longitudinal scalar modes (see Fig.~\ref{fig:Polarizations}).
 
In addition, it will be possible to test additional polarizations with continuous GW sources such as pulsars \cite{Isi:2017equ}. No signal has so far been detected \cite{Aasi:2015rar,Abbott:2017ylp}, although only the first run has been analyzed because of the costly computational analysis. 
 
Finally, observing the stochastic backgrounds of GWs can probe as well non-GR polarizations. Such background is composed of individually unresolved sources. Since the signal is received from different points in the sky in a continuous manner, it allows a direct measurement of the polarization from the spectral shape of the background \cite{Callister:2017ocg}. No stochastic GW background has been detected yet, placing limits on the stochastic background from tensor, vector and scalar polarizations \cite{Abbott:2018utx}.

 \subsection{Gravitational wave oscillations}
 \label{sec:GWoscillations}
 
\begin{figure}
\centering 
 \includegraphics[width=0.99\columnwidth]{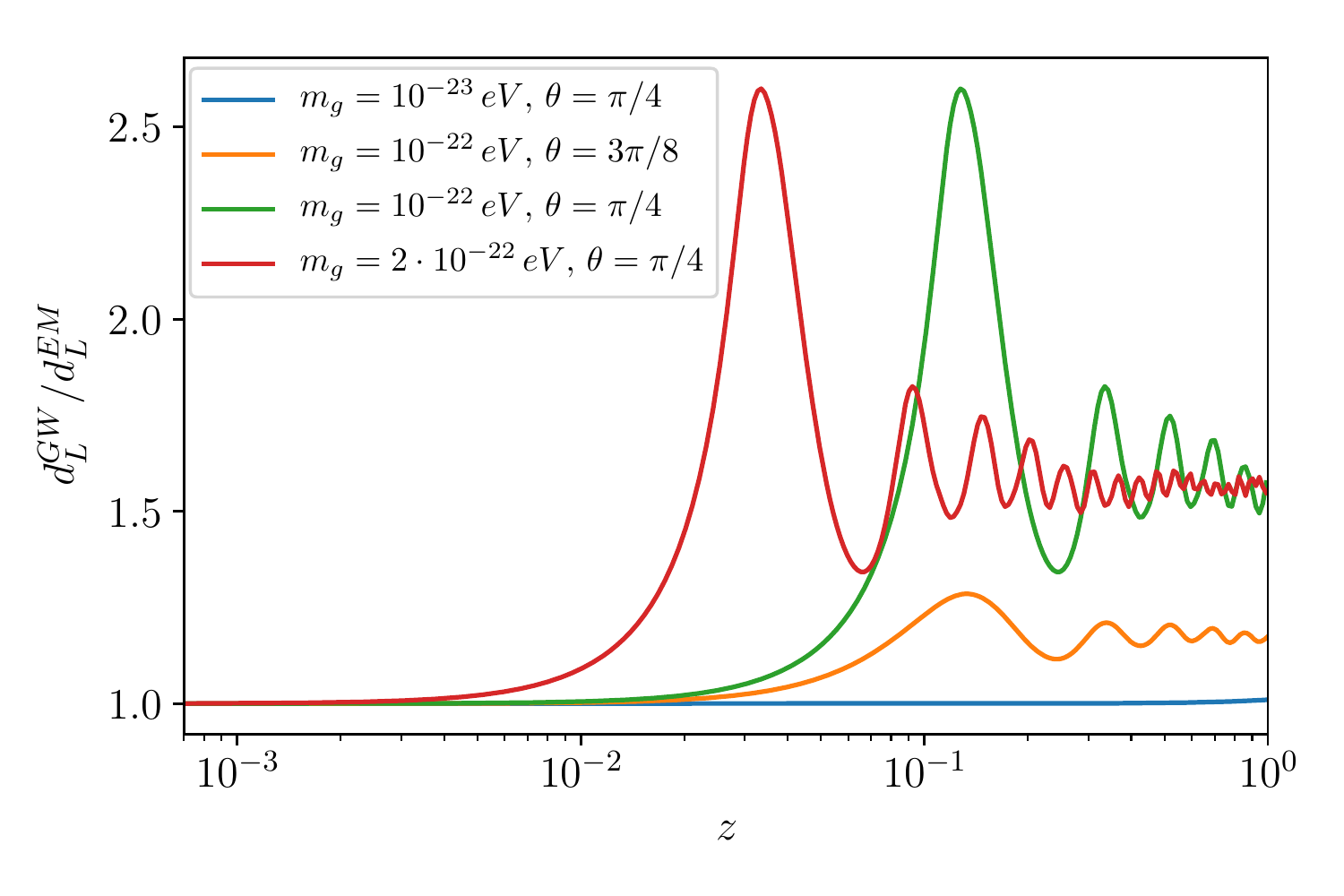}
 \caption{Modulation of the GW luminosity distance due to GW oscillations in bigravity for different graviton masses $m_g$ and mixing angles $\theta$, cf. (\ref{eq:bigravity}). This plot is adapted from the results of \cite{Max:2017flc}.}
 \label{fig:GWoscillations}
\end{figure}
 
Interestingly, these extra modes might mix with the GR polarizations $h_{+,\times}$. Over cosmological backgrounds, tensor polarizations can only mix with other tensor modes by symmetry arguments. The simplest example of a theory with two metric perturbations is bigravity. In analogy with neutrino oscillations, the difference between the mass and propagation eigenstates in bigravity leads to GW oscillations \cite{Narikawa:2014fua,Max:2017flc,Max:2017kdc}. Assuming that GWs are emitted as in GR, these oscillations during the propagation introduce a modulation of the GW amplitude. Thus, depending on the mixing angle and the mass of the graviton, there will be oscillations in the GW luminosity distance as a function of redshift. We plot different examples in Fig.~\ref{fig:GWoscillations}. Present ground-based detectors are sensitive to masses $m_g\sim10^{-22}\,\text{eV}$. The mixing is maximized at an angle $\theta=\pi/4$ (recall (\ref{eq:bigravity})). In principle, for several multi-messenger events at different redshifts one could trace these oscillations \cite{Max:2017flc}. Moreover, with space-based detectors like LISA one could reach a thousand times smaller masses. Interestingly, since both perturbations travel at different speeds due to the mass term of one of them, it is possible that they decohere ending traveling independently and arriving at different times. GW detectors will then see an echo signal. This allows to further constraint the parameter space of bigravity \cite{Max:2017kdc}.
 
Finally, we should stress that GW oscillations are not a unique property of bigravity. For instance, gravity theories with gauge fields in an SU(2) group have effectively two metric perturbations as well, leading to the same phenomena \cite{Caldwell:2016sut,Caldwell:2018feo}. This can happen in different classes of vector-tensor DE models too \cite{BeltranJimenez:2018ymu}.
  
\section{Theoretical implications}
\label{sec:TheoreticalImplications}

Present GW observations place severe limits on deviations from GR. Among the different constraints, the most stringent ones are the propagation speed equal to the speed of light and the absence of emission of additional polarizations. The key question is then
\begin{displayquote}
\emph{within the set of theories passing present tests, what interesting phenomenology is still possible?}
\end{displayquote}  
Of course, we do not have a complete answer to this question. In the following, we survey different proposals of viable theories and highlight some lessons we have learned in light of current bounds. 

\subsection{$c_g=c$} \label{sec:cg_consequences}

 Before considering which theories are compatible with present constraint on the speed of GWs, it is important to discuss how far reaching this new measurement is. The first thing to note is that due to the closeness of the BNS, the constraint only applies basically to present time. Therefore, one could envision a situation in which the speed of GWs was different from the speed of light at early times but due to the cosmological evolution at present time $c_{g}(z=0)=c$. However, one should be careful about this argument for several reasons. First, the level of precision of $c_g/c-1$ requires the cosmological evolution to be tuned at the level of 1 part in $10^{15}$. One way around this argument is to have $c_{g}(z=0)=c$ as a late time cosmological attractor. An example of this is Doppelg\"anger DE \cite{Amendola:2018ltt}, where a coupling between DM and DE allowed for this attractor to exist. Still, if the derivative couplings to the curvature leading to the anomalous speed remain present in the action, there are reasons to worry \cite{Ezquiaga:2017ekz}. This is because although the cosmological evolution might lead to the precise cancellation of the dangerous terms, there will be deviations from the cosmological background along the path of the GWs towards the detector, for instance, when they cross the Milky Way. 

A second remark is that constraints in the dispersion relation only apply to the characteristic wave numbers of the compact binary systems detected so far. These modes are characterized by $k_\text{gw}\gg H_0$. As a consequence, in a phenomenological approach, one could envision modifications of the dispersion relation only arising at cosmological scales \cite{Battye:2018ssx}, for instance
\be
\omega^2(k)=c_g^2k^2\lp1+\sum_{n}c_n\lp \frac{aH}{k}\rp^{n}\rp\,.
\ee
This could, in principle, lead to modified gravity effects at large scales not affecting present GW constraints. However, in practice, only theories with non-local couplings or higher derivative interactions with ghost degrees of freedom are known to have this dispersion relation. It would be interesting to study in depth the theoretical framework allowing for this modified propagation.

Related to this point, one should note that the frequency of GW170817 was close to the typical strong coupling scale of the EFT of DE $\Lambda_\text{strong}\sim (M_\text{pl}H_0^2)^{1/3}\sim260\,\text{Hz}$. If the cutoff of the theory is of the order of the strong coupling scale $M_\text{cutoff}\sim\Lambda_\text{strong}$, as it is usually assumed, higher dimensional operators might modify the dispersion relation although one would not expect that they conspire to completely cancel the anomalous speed at the level of $\mathcal{O}(10^{-15})$ \cite{Creminelli:2017sry}. 
In the case in which the cutoff scale is parametrically smaller, $M_\text{cutoff}\ll\Lambda_\text{strong}$, the situation could be different \cite{deRham:2018red}.  
Theories with a Lorentz-invariant ultra-violet (UV) completion are presumed to have luminal GW propagation. Therefore, one would expect higher dimensional operators to erase any anomalous speed beyond the cutoff scale, which in this case might already happen in the LIGO band. However, the speed of GWs cannot be computed beyond $M_\text{cutoff}$ if the UV completion is unknown.
In any case, the hypothesis that higher dimensional operators render $c_g(k_{_\text{LIGO}})=c$ could be tested detecting GWs at different frequencies, for example with LISA (see Fig. \ref{fig:Sensitivity}). This might give us valuable information about the cutoff scale of the effective theory of DE.

Another lesson from GW170817, as it was discussed in Sec. \ref{sec:AnomalousSpeed}, is that the effective metric for GWs is proportional to the effective metric for EM radiation. In other words, the GW-cone and the light-cone are the same. This fact suggests two ways to construct theories with $c_g=c$ \cite{Ezquiaga:2017ekz}. On the one hand, one could start with a theory in which GWs propagate at the speed of light and apply a conformal transformation $\tilde g_{\mu\nu} = \Omega^2(\phi,X) g^{_\text{B}}_{\mu\nu}$ to the gravity sector \footnote{Note that if the field redefinition was applied to the whole action, the transformed theory will not lead any new physics, being completely equivalent to the original one.}. Then, one would automatically arrive to a theory with $c_g=c$. In the case of scalar-tensor gravity, if one applies this recipe to GR, one arrives at the kinetic conformal theory (\ref{eq:kin_conf_theory}), which was the first extension beyond Horndeski \cite{Zumalacarregui:2013pma}. On the other hand, one could begin with a theory with $c_g\neq c$ and apply a disformal transformation \cite{Bekenstein:1992pj},
\be 
\tilde g_{\mu\nu} = \Omega^2(\phi,X) g^{_\text{B}}_{\mu\nu}+\mathcal{D}(\phi,X)\phi_{,\mu}\phi_{,\nu}\,,
\ee
engineered to compensate the anomalous speed. This is because the term $\mathcal{D}$ is not proportional to the metric and can modify the causal structure, unlike the conformal term $\Omega^2$. This is clear when computing how the speed of GWs would transform \cite{Ezquiaga:2017ekz}
\be
\tilde c_g^2=\frac{c_g^2(\tilde X)}{1+2\tilde X \mathcal{D}}\,, \label{eq:disfspeed}
\ee
where $c_g$ is the speed of tensors of the original gravity theory and $-2\tilde X=\tilde g^{\mu\nu}\phi_{,\mu}\phi_{,\nu}$.\footnote{This result can be proven explicitly using the full disformal transformation of Horndeski theory presented in Ref. \cite{Ezquiaga:2017ner}.} In this case, starting with Horndeski theory, one would arrive at the subclass of GLPV theory \cite{Gleyzes:2014dya} characterized by $c_g=c$. In terms of the free functions in the action (\ref{eq:glpv4}), one needs to impose $B_4=G_{4,X}/X$. Satisfying this constraint, concrete DE models have been proposed \cite{Kase:2018iwp}. In the context of DHOST theories, this constraint implies $A_1=0$. Something to note is that after GW170817, DHOST has the same number of free functions in the action as Horndeski had before the constraint on the speed of GWs, i.e. four free functions of $\phi$ and $X$ that could be counted as $K(\phi,X)$, $G_{3}(\phi,X)$, $G_{4}(\phi,X)$ (with $B_4=G_{4,X}/X$) plus the conformal redefinition $\Omega^2(\phi,X)$ (cf. (\ref{eq:LH2}-\ref{eq:LH4},\ref{eq:glpv4},\ref{eq:kin_conf_theory})). 

One may worry whether the conditions for $c_g=c$ are stable under quantum corrections. If they are not, one would need to tune the GW speed order by order in perturbation theory. Using the results of \cite{Pirtskhalava:2015nla,Luty:2003vm} linking the properties of Horndeski with those of Galileons, it was argued in Ref. \cite{Creminelli:2017sry} that the quantum corrections to the EFT coefficients are negligible, $\mathcal{O}(10^{-40})$, even compared to the $10^{-15}$ constraint in $c_g/c-1$. Thus, the tree-level condition is not modified (see also \cite{Santoni:2018rrx} where the same conclusion is derived analysing higher derivative EFTs).

Within the scalar-tensor theories compatible with the constraint on the speed of GWs, there have been extensive efforts to explore interesting phenomenology. One immediate question is whether the survival theories can provide accelerated expansions at late times without a cosmological constant as covariant Galileons were providing. This possibility was investigated in the context of DHOST theory \cite{Crisostomi:2017pjs}. It was found that indeed there are scaling solutions with a late time de Sitter behavior. Still, a full comparison with present cosmological observations is missing due to the lack of appropriate Boltzmann codes for these higher-derivative theories. 

Another attractive feature of Horndeski gravity was the possibility to have self-tuning solutions \cite{Charmousis:2011bf,Charmousis:2011ea}. This was an attempt to solve the cosmological constant problem by counterbalancing the large bare vacuum energy with the energy momentum of the scalar field. However, Fab Four models realizing this behavior predict an anomalous GW speed. Now, beyond Horndeski models with $c_g=c$ could also exhibit self-tuning. Indeed, an infinite set of self-tuning models compatible with GW170817 were found in \cite{Babichev:2017lmw}. Again, a detailed cosmological analysis is left for future work.

In the realm of Horndeski gravity, one could search for other definite predictions. In addition to the condition on the speed of GWs ($\alpha_{_\text{T}}=0$) one could impose that the gravitational strength coupling to matter is the same as the one to light, $G_\text{matter}=G_\text{light}$ (or $\alpha_{_\text{B}}=-2\alpha_{\text{M}}$). This model, named no slip gravity \cite{Linder:2018jil}, has the property of predicting that gravity should be weaker than GR in the late universe. This could be tested with growth of structure observations in the next generation galaxy redshift surveys.

\subsection{Compact objects}
\label{sec:CompactObjects}

Present observations severely constrain deviations from GR at small scales. Screening mechanisms are thus needed to surpass these bounds \cite{deRham:2012fw,Chu:2012kz,deRham:2012fg,Barausse:2015wia}. Modified gravity theories can display different types of screening mechanisms (see reviews \cite{Brax:2013ida,Joyce:2014kja}). For theories with derivative interactions this is achieved with the Vainshtein mechanism, which screens the fifth force when the local curvature is larger than a given threshold. Such mechanism has been extensively studied for Horndeski theory \cite{Kimura:2011dc,Narikawa:2013pjr,Koyama:2013paa}. For theories beyond Horndeski of the GLPV class the screening works similarly outside the source, but there is a breaking of Vainshtein screening inside matter \cite{Kobayashi:2014ida}. This suggests using astrophysical systems, such as neutron stars, to test these theories \cite{Koyama:2015oma,Babichev:2016jom}.

The question then is whether the viable scalar-tensor theories (in light of GW tests) can display a successful screening and if there are any observational signatures to test them. This was addressed by different groups soon after the announcement of GW170817 \cite{Langlois:2017dyl,Crisostomi:2017lbg,Dima:2017pwp}. One should note that many models in which the breaking of Vainshtein screening was studied previously are incompatible with $c_g=c$. Still, these recent analyses show that within DHOST theories satisfying the constraint in the speed of GWs, screening is effective outside non-relativistic bodies, but there could be a breaking inside matter as well. This deviation from GR inside compact bodies is only predicted for theories beyond Horndeski. Comparing with the previous GLPV analysis, the weakening of Vainshtein screening inside matter in DHOST theories has a different form with an additional term not present before. 

Moreover, the emission of additional polarizations is highly constrained as well. 
Depending on the gravity theory, compact objects might emit extra radiation (see \cite{Herdeiro:2015waa} for a review on no-hair theorems).
An interesting question is if cosmologically relevant theories compatible with the bound on the speed of GWs can exhibit scalar hair in the black-hole solutions. In \cite{Tattersall:2018map} it was found under these conditions only very little or no scalar hair is possible. Analysis of black-hole solutions including screening effects have not been studied so far. Strong field effects are yet possible in theories not aimed at explaining cosmology, for instance spontaneous scalarization in neutron stars \cite{Damour:1993hw} or even in black-holes \cite{Silva:2017uqg,Antoniou:2017acq,Doneva:2017bvd} (see more details in the extensive review \cite{Barack:2018yly}). However, one should note that this kind of solutions may also induce an anomalous propagation speed due to the spatial scalar field profile~\cite{Papallo:2015rna}. Possible constraints from this effect should be investigated further.

\section{Conclusions and outlook}
\label{sec:Conclusions}

Gravitational wave astronomy has opened a new window to test gravity and dark energy. Multi-messenger probes are specially promising for this task. The first detection of GWs from a binary neutron star merger, GW170817, was followed up by several EM counterparts. This has provided an independent, standard siren measurement of the Hubble constant $H_0$. Moreover, GW170817 already constrains large classes of DE models. In particular, the bound on the speed of GWs was significantly strong. Other multi-messenger tests of DE are possible, such as probing the GW luminosity distance or searching for additional polarizations. These tests will become more relevant in the future when more events will be available. Still, there remain important challenges in this GW program to probe DE.

From the observational side, it will be crucial to achieve a global synergy in the quest of multi-messenger GW astronomy. On the one hand, GW detectors will have to improve their sensitivity and enlarge the network to detect more events and localize them better. On the other hand, observatories around the world should be available to follow-up triggers. Moreover, improved galaxy catalogues might be necessary to maximize the chances of localization. Lastly, cross-correlations between GWs and other cosmological probes might be an interesting endeavor.

From the theoretical side, the main challenge will be to analyze the GW propagation over non-cosmological backgrounds, understanding the possible interplay of additional polarizations. This will be relevant for instance for GWs traveling through a screened region. Furthermore, degeneracies between modified gravity predictions and astrophysical properties should be studied in more detail.  For example, possible signatures of phenomenology beyond GR in neutron stars could possibly be the same as modifications of the equation of state.  

Altogether, the future of multi-messenger GW astronomy appears promising. In the coming years standard sirens will be routinely detected and we will be able to apply the different GW tests of gravity to a much higher precision. 
The new techniques brought by GW astronomy will bring us closer to unveil the nature of dark energy.

\begin{acknowledgments}
We are grateful to E. Bellini, J. Beltr\'an, J.L. Bernal, D. Bettoni, D. Blas, L. Heisenberg, G. Horndeski, J. Smirnov, G. Tasinato and F. Vernizzi for comments on the manuscript. 
We would like to thank also the authors of \cite{Max:2017flc,Max:2017kdc} for providing us the tools to adapt their plots, and the authors of \cite{Renk:2017rzu} and \cite{Belgacem:2017cqo} for the permission to reproduce their figures.
JME is supported by the Spanish FPU Grant No. FPU14/01618, the Research Project FPA2015-68048-03-3P (MINECO-FEDER) and the Centro de Excelencia Severo Ochoa Program SEV-2016-0597. He thanks CERN Theory Division for hospitality during his stay there and the FPU program for financial support.
MZ is supported by the Marie Sklodowska-Curie Global Fellowship Project NLO-CO. He thanks the Instituto de F\'isica Fundamental IFF-CSIC for their hospitality during the completion of this work.
We acknowledge the use of the \texttt{hi\_class} code (\url{www.hiclass-code.net}) \cite{Zumalacarregui:2016pph}.
This research has made use of data, software and/or web tools obtained from the LIGO Open Science Center (https://losc.ligo.org), a service of LIGO Laboratory, the LIGO Scientific Collaboration and the Virgo Collaboration. LIGO is funded by the U.S. National Science Foundation. Virgo is funded by the French Centre National de Recherche Scientifique (CNRS), the Italian Istituto Nazionale della Fisica Nucleare (INFN) and the Dutch Nikhef, with contributions by Polish and Hungarian institutes.
The first version of this work appeared as a preprint \cite{Ezquiaga:2018btd}.
\end{acknowledgments}

\bibliography{gw_refs}
\end{document}